%% file: main.tex
\documentclass[pdflatex,sn-mathphys-num,iicol]{sn-jnl}

\usepackage{amsmath}
\usepackage[T1]{fontenc}
\usepackage{hyperref}
\usepackage{siunitx}
\usepackage{graphicx}
\usepackage{xurl} 
\usepackage{soul}
\usepackage[mode=buildnew]{standalone}%

\usepackage{subcaption}
\hypersetup{colorlinks = true}
\input{abrev} 
\usepackage{etoolbox}
\patchcmd{\abstract}{\centerline{\large\bfseries Abstract}}{\centerline{\large\bfseries Abstract}}{}{}

\begin{document}

\title{Reconstruction of atmospheric neutrinos in DUNE's horizontal-drift far-detector module}
\input{authorlist/export_xml_authors_2025-07-22_EPJC}


\abstract{
This paper reports on the capabilities in reconstructing and identifying atmospheric neutrino interactions in one of the Deep Underground Neutrino Experiment's (DUNE) far detector modules, a liquid argon time projection chamber (LArTPC) with horizontal drift (FD-HD) of ionization electrons. The reconstruction is based upon the workflow developed for DUNE's long-baseline oscillation analysis, with some necessary machine-learning models' retraining and the addition of features relevant only to atmospheric neutrinos such as the neutrino direction reconstruction. Where relevant, the impact of the detection of the charged particles of the hadronic system is emphasized, and comparisons are carried out between the case when lepton-only information is considered in the reconstruction (as is the case for many neutrino oscillation experiments), versus when all particles identified in the LArTPC were included. Three neutrino direction reconstruction methods have been developed and studied for the atmospheric analyses: using lepton-only information, using all reconstructed particles, and using only correlations from reconstructed hits. The results indicate that incorporating more than just lepton information significantly improves the resolution of both neutrino direction and energy reconstruction. 
The angle reconstruction algorithms developed in this work result in no strong dependence on particle direction for reconstruction efficiencies or neutrino flavor identification. This comprehensive review of the reconstruction of atmospheric neutrinos in DUNE's FD-HD LArTPC is the first step towards developing a first neutrino oscillation sensitivity analysis, which will ready DUNE for its first measurements.

}

\maketitle


\tableofcontents

\clearpage
\input{Intro}
\input{AnaSetup}
\input{AtmReco}
\input{Summary}
\input{Outlook}

\section{Acknowledgements}
\input{aknowlege}

\bibliography{references}
\end{document}

%% file: abrev.tex
\newcommand{\tetay}{$\theta^{\rm{true}}_{y}$}

\newcommand{\be}{\begin{eqnarray}}
\newcommand{\ee}{\end{eqnarray}}

\def\BE{\begin{equation}}
\def\EE{\end{equation}}

\newcommand\dedx{$\frac{dE}{dx}$\,}
\newcommand{\dmTwoThreeSq    }{$\Delta m^{\text{2}}_{\text{32}}$}

\newcommand{\thTwoThree   }{$\theta_{\text{23}}$}
\newcommand{\dCP       }{$\delta_{\text{CP}}$}

\newcommand{\numu   }{$\nu_\mu$}
\newcommand{\nue    }{$\nu_e$}
\newcommand{\nutau  }{$\nu_\tau$}

\newcommand{\numubar}{$\bar{\nu}_\mu$}
\newcommand{\nuebar }{$\bar{\nu}_e$}

\newcommand\ie {{\it i.e.}}
\newcommand\eg {{\it e.g.}}
\newcommand\etc{{\it etc.}}

\newcommand\vs {{\it vs.}}

\newcommand{\mcs}{MCS}
\newcommand{\mcsname}{Multiple Coulomb Scattering}
\newcommand{\erecollhd}{LLHD}
\newcommand{\erecochi}{Chi2}

\newcommand{\csda}{CSDA}

\def\GeV{\,{\rm GeV}}
\def\MeV{\,{\rm MeV}}

\newcommand{\Ekin}{$E_{\rm kin}$}  
\newcommand{\Ekintrue}{$E_{\rm kin}^{\rm true}$} 
\newcommand{\Ekinreco}{$E_{\rm kin}^{\rm reco}$} 
\newcommand{\Ehadreco}{$E_{\rm had}^{\rm reco}$} 
\newcommand{\Elepreco}{$E_{\rm lep}^{\rm reco}$} 
 
\newcommand{\Etruereco}{$E_{\rm had}^{\rm true}$} 

\newcommand{\Enu}{$E_{\nu}$} 
\newcommand{\Enutrue}{$E_{\nu}^{\rm true}$} 
\newcommand{\Enureco}{$E_{\nu}^{\rm reco}$} 
\newcommand{\Eereco}{$E_{e}^{\rm reco}$}
\newcommand{\Emureco}{$E_{\mu}^{\rm reco}$}

%% file: authorlist/export_xml_authors_2025-07-22_EPJC.tex
%

\author[114]{S.~Abbaslu}
\author[81]{F.~Abd Alrahman}
\author[35]{A.~Abed Abud}
\author[35]{R.~Acciarri}
\author[193]{L.~P.~Accorsi}
\author[12]{M.~A.~Acero}
\author[193]{M.~R.~Adames}
\author[72]{G.~Adamov}
\author[66]{M.~Adamowski}
\author[30]{C.~Adriano}

% ...
\affil[12]{Universidad del Atl\'antico, Barranquilla, Atl\'antico, Colombia}
% ...
\affil[30]{Universidade Estadual de Campinas, Campinas - SP, 13083-970, Brazil}
% ...
\affil[35]{CERN, The European Organization for Nuclear Research, 1211 Meyrin, Switzerland}
% ...
\affil[66]{Fermi National Accelerator Laboratory, Batavia, IL 60510, USA}
% ...
\affil[72]{Georgian Technical University, Tbilisi, Georgia}
% ...
\affil[81]{University of Houston, Houston, TX 77204, USA}
% ...
\affil[114]{Institute for Research in Fundamental Sciences, Tehran, Iran}
% ...
\affil[193]{Universidade Tecnol\'ogica Federal do Paran\'a, Curitiba, Brazil}
% ...

%% file: Intro.tex
\section{Introduction}
\label{intro}


The Deep Underground Neutrino Experiment (DUNE) is under construction in the U.S.~\cite{dunefdintro}. Its near detector (ND) is to be located at Fermilab in Illinois near the high-intensity LNBF beam target generated from a megawatt-class proton accelerator, while its far detector (FD) is hosted at a distance (or baseline) $L\sim$1300\,km away and $\sim$1.5\,km underground at Sanford Underground Research Facility (SURF) in South Dakota. Together, the ND and FD will enable a broad exploration of neutrinos and the physics laws that govern them~\cite{dunephys}. DUNE's primary science mission is to test the three-flavor paradigm of neutrino physics, determine the unknown neutrino mass ordering, make precise measurement of neutrino mixing parameters and determine the degree of charge-parity violation in the lepton sector, a key ingredient in certain theories seeking to resolve the origin of the baryon abundance of the universe~\cite{dunepotential,FUKUGITA198645, DUNE:2020zfm,DUNE:2020fgq,DUNE:2021mtg}. 

The success of this program are the massive time projection chamber (TPC) modules of the FD, filled with a fiducial mass of 40 kilotons of liquid argon (LAr) ~\cite{dunevd,dunehd}. These \mbox{LArTPCs} will enable neutriono interactions to be resolved with high precision through the processing and reconstruction of ionizing activity down to order-millimeter length scales. It is expected that the first two modules of the FD (filled with 17 kilotons of LAr each) will be installed and ready to collect data a few years before the LBNF beam comes online. This presents the opportunity to start the physics program using natural sources of neutrinos, in particular neutrinos from secondary decays of particles produced by cosmic ray interactions in Earth's atmosphere, \ie, atmospheric neutrinos. 

DUNE has sensitivity to oscillation parameters determined using atmospheric neutrinos~\cite{dunephys}. In particular, measurements of \dmTwoThreeSq\ can be carried out~\cite{Ternes:2019sak} to address the unknown neutrino mass ordering, along with the \thTwoThree\ mixing angle to resolve its octant degeneracy ambiguity~\cite{dunephys}, and finally, a complementary beam-independent measurement of the \dCP\ charge parity violating phase using sub-\GeV\ atmospheric neutrinos~\cite{Kelly:2019itm}. Moreover, with $\mathcal{O}(10^3)$ \nue\ and \numu\ atmospheric interactions expected per year in each of the FD modules~\cite{dunephys}, and the significant energy overlap with beam neutrino interactions, a high-level of complementarity is expected between measurements performed across both sources. The reconstruction capabilities of LArTPC technology allows for particle detection and identification beyond only leptons; with a deep underground location protecting the large mass FDs from cosmogenic backgrounds, it is expected that high quality, high statistics measurements of atmospheric neutrinos are possible in DUNE. This will open opportunities for beyond Standard Model (BSM) physics studies~\cite{Super-Kamiokande:2014exs,IceCube:2017qyp}, including sterile neutrino searches~\cite{Super-Kamiokande:2014ndf}, non-standard interactions~\cite{Chatterjee:2014gxa}, and mass-varying neutrinos~\cite{Super-Kamiokande:2008nzt}. Critical interrogation and correlation of atmospheric neutrino and cosmogenic activities can also aid in precision background predictions relevant for baryon number violation and (boosted) dark matter searches in the DUNE FDs. Overall, it is expected for DUNE measurements to complement and advance the atmospheric neutrino program carried out by present (IceCube~\cite{IceCube}, KM3NeT~\cite{KM3NeT}, Super-Kamiokande~\cite{SK}, \etc) and future (Hyper-Kamiokande~\cite{HK}) neutrino experiments. 

In general, there are clear advantages in exploiting atmospheric neutrinos: i) they are available at all times; ii) the oscillated flux contains all flavors of neutrinos and antineutrinos; iii) their energy \Enu\ spans a wide range (0.1 -- 100\GeV) corresponding to roughly five orders of magnitude in $L/$\Enu\ ($\sim$ 1 -- \SI{6e5}{km/\GeV}), broadening the reach of a neutrino physics program (\eg, the BSM studies); iv) they can serve to break beam-only oscillation degeneracies and increase overall sensitivities via a future joint analysis, or act as a cross check method of common reconstruction techniques utilizing the same detectors; and v) they can be used for commissioning of the detectors in preparation for the beam neutrinos. Compared to their beamline counterparts, atmospheric neutrinos bring some complications: i) their flux is not known to the precision expected for beam neutrinos, depending on theoretical modeling of data from other experiments; ii) their direction is not known, which makes it necessary for the neutrino reconstruction to be understood and optimized for all incident angles; iii) the oscillation baseline (\ie, the distance from where they are produced to the detectors) and the Earth's structure, which they pass through, is less precisely known and model dependent, respectively; iv) their interaction time ($t_0$) in the detector is not known, and requires additional measurements to determine the time when the ionization electrons begin to drift; v) the reconstruction of their wide range of energies is highly challenging, as the interaction processes and associated topologies vary widely as a function of incident neutrino energy; and, finally, vi) it is harder to distinguish neutrino from antineutrino in the atmospheric flux, which in the case of the beam can be achieved by using magnet-focusing horns to sign-select the mesons generating the (anti)neutrino beam.

Neutrinos' energy, direction, and flavor are reconstructed indirectly in a detector by identifying and reconstructing the aftermath of its interaction with the active medium of the detector. The accuracy of a neutrino interaction reconstruction directly impacts associated physics measurements, which depend inherently upon how precisely the position of the neutrino interaction vertex is determined, and on how well the energy, direction, and particle identification (PID) reconstruction algorithms perform for all daughter particles stemming from this vertex. This paper focuses on the deployment and validation of a complete set of reconstruction tools for atmospheric neutrinos in DUNE, which is an important stepping stone for any future physics analysis. Here, the performance of reconstruction algorithms will be presented for atmospheric neutrinos with energies between \SI{100}{MeV} and \SI{10}{GeV}, using a simulated DUNE FD LArTPC module. The upper limit is guided by the containment boundaries imposed by the reduced detector geometry used for this study. Particle reconstruction efficiencies and resolutions are evaluated, and their energy and direction dependence is reported.

The paper is structured as follows. First, a description will be given for the detector simulation setup, the reconstruction software, and simulated data sample used. Then, reconstruction algorithms relevant for the neutrino vertex, angle, and energy reconstruction will be presented, with performance assessments. The particle reconstruction and tagging performance, as well as a neutrino flavor identification method will also be discussed. Finally, a summary of the overall neutrino interaction reconstruction performance will be given, with an overview of the most promising future developments, and closing with a discussion of this status in context of other atmospheric neutrino detectors.

%% file: AnaSetup.tex
\section{Analysis setup}
\label{sec:anaSetup}

\subsection{Sample generation and reconstruction}

The first two FD modules that will start taking data will both be
LArTPCs~\cite{dunefdintro}. However, their design differs, with most changes stemming from the choice of the cathode position. For the vertical-drift (VD) design, the cathode is suspended horizontally mid-way between the top and bottom of the module~\cite{dunevd}. For the horizontal-drift (HD) design, the cathode planes are positioned vertically~\cite{dunehd}. An electric field is applied to the active LAr volume, with different drift directions (vertical \vs\ horizontal) and distances (6.5 \vs\ \SI{3.6}{m}). Distinct charge readout technologies (copper strips printed on large circuit readout planes, CRPs, vs wires wound in large anode planes assemblies, APAs) are used for the two modules.
However, the Pandora Software Development Kit (SDK)~\cite{pandora_uboone, pandora_sdk}, which currently serves as the main package used in DUNE to provide pattern recognition algorithms to reconstruct the neutrino interaction, is agnostic to the detector differences, and specific orientations between incoming neutrino and detector drift directions were found to have mild effects. As a result, a comparable atmospheric neutrino reconstruction performance overall is expected for horizontal- and vertical-drift detector configurations, allowing this article to focus exclusively on the Horizontal-Drift detector.


Charged particles passing through LAr ionize and excite the argon atoms producing ionization electrons and scintillation photons. These photons are detected by DUNE’s Photon Detection System (PDS) and reconstructed as a collection of flashes~\cite{dunephys}.
It was checked that every selected atmospheric neutrino interaction for which results are reported has an associated reconstructed PDS flash. In addition, charge-light mismatches, which incorrectly combine atmospheric neutrino TPC information with reconstructed PDS flashes produced by unrelated, and lower-energy, radiogenic-induced energy deposits, are also expected to be small in this energy range. Therefore, $t_0$ and drift distance determination is assumed to be perfect in the following, and no further use of the PDS information is made in this paper.


 \begin{figure*}[tb]
    \centering
    \includegraphics[width=0.7\linewidth]{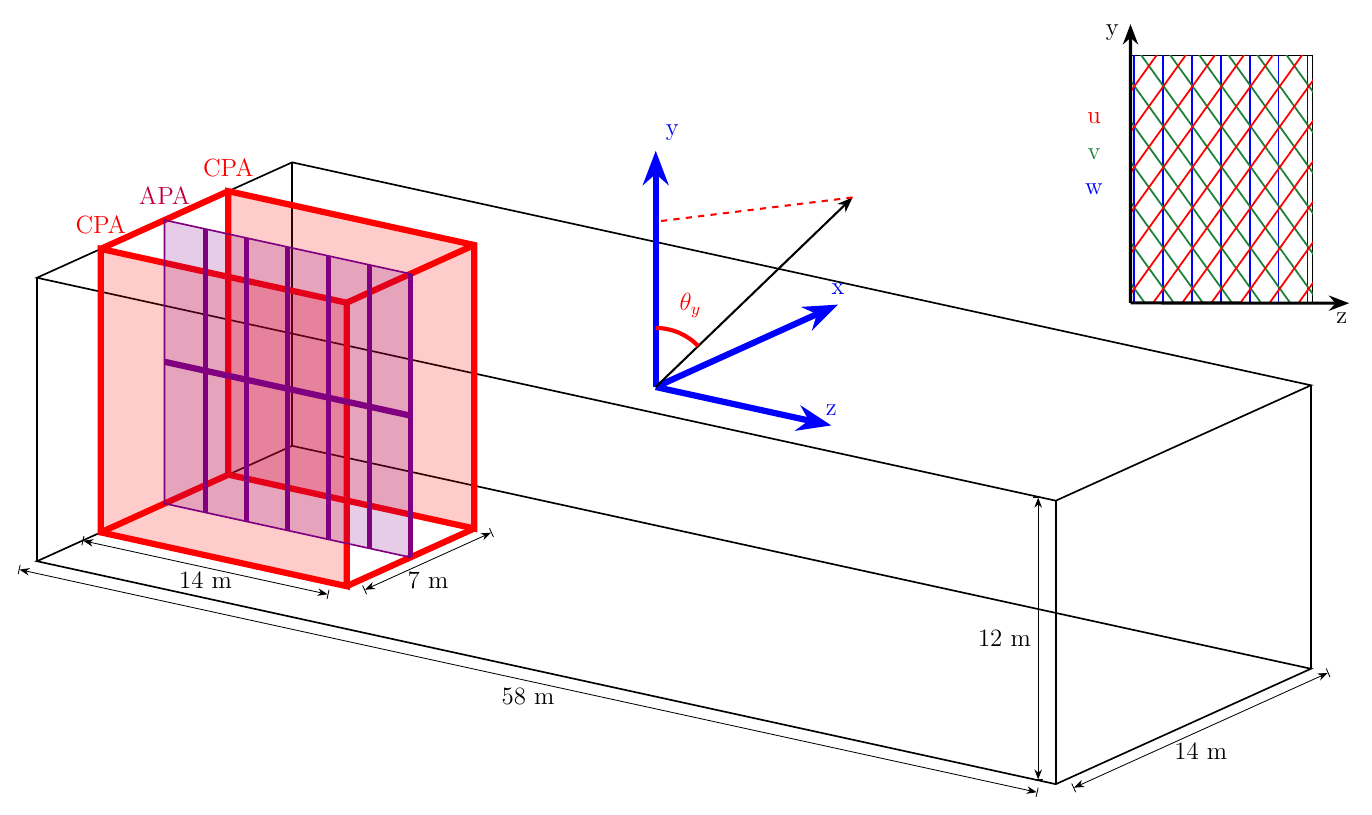}
    \caption{At-scale depiction of the `DUNE HD 1x2x6' geometry (red) within the full HD module geometry (black). The plane of APAs (purple) is flanked by the two CRP planes (light red). Also shown, the wires layout of the planes $u$, $v$, and $w$, together with the right-handed coordinate system used in which the zenith angle ($\theta_y$) is marked.}
    \label{fig:geometry}
\end{figure*}

The ionization electrons are carried by the electric field to the APA, where three layers ($u$, $v$, and $w$) of active wires (vertically oriented for the $w$ plane, and at $\pm 35.7^{\circ}$ to the vertical for $u$ and $v$ planes), form a regular grid for collection and induction of ionization charge readout. The relative voltage between the planes is chosen to ensure all but the final layer ($w$) are transparent to the drifting electrons. The right-handed Cartesian coordinate system used to describe the detector geometry, depicted in Fig.~\ref{fig:geometry}, has $x$ as the drift axis, $y$ as the upwards vertical direction, and $z$ the remaining orthogonal direction. While in the case of beam neutrinos $z$ is the primary direction, for atmospheric neutrinos the zenith angle \tetay{} is used throughout the paper when reporting reconstruction performance dependence on neutrino direction. This choice is motivated by \tetay{} being the only relevant angle for sensitivity studies with atmospheric neutrinos. For each readout plane, localized charge deposits (\ie, hits) are identified from the collected or induced charge waveforms, with each plane ($u$, $v$, $w$) providing a 2D `view' (U, V, W, respectively) of particle interactions in the LArTPC.  

The three sparse lists of 2D hits (hit wire and time) are provided as input to Pandora's large number of decoupled algorithms to develop the reconstruction from hits to particles, with each algorithm designed to address a specific aspect of pattern recognition. DUNE's software~\cite{pandora_protodune} builds directly on the Pandora developments completed for another LArTPC detector, MicroBooNE~\cite{pandora_uboone}. The most important output of the pattern recognition is the list of Particle Flow Particles (PFPs). Each PFP is associated with: i) a list of 2D clusters, which group together the relevant hits from each readout plane based on their topological associations; ii) a set of reconstructed 3D positions, which combine the 2D information from the three planes for each cluster; and iii) a reconstructed vertex position, defining the PFP's interaction point. As part of the PFP reconstruction, Pandora also provides a score describing how likely the PFP is to be a track-like or a shower-like particle, \ie, whether its energy depositions in the detector are more like that of a muon, a pion, or of a proton (linear pattern) or like that of an electron or a photon (spray pattern). The PFPs are placed in a hierarchy, corresponding to their identified parent-child order, \ie, which `child' particle was produced during the decay or interaction of a `parent' particle. A neutrino PFP will be the primordial parent for a neutrino interaction. An identification of each particle, as well as the flavor determination for the neutrino, can also be made, and some results will be discussed in this paper.

Hereafter, a simulated event is defined as a single readout window of \SI{2.25}{ms}—spanning the full ionization electron drift time—aligned such that the neutrino interaction occurs at $t_0 = 0$. Given the rock overburden, the rate of cosmic ray  muons is expected to be approximately $\mathcal{O}(10^{-4})$ per readout window in the full volume of a horizontal drift module, and therefore no cosmics are simulated. Neutrino interactions are generated using GENIE v3.04.00~\cite{Andreopoulos:2009rq, genie} with the Liquid Argon Experiment tune AR23\_20i\_00\_000, the propagation of particles and their interaction within the detector are simulated by Geant4 v10.6.p01f~\cite{geant, geant_update1, geant_update2} with the QGSP\_BERT physics list, and the detector response is simulated using LArSoft v09\_81\_00~\cite{larsoft, larsoftWeb}. The physics parameters used for the LArTPC simulation have been informed by the ProtoDUNE reported performance and measurements~\cite{DUNE:2020cqd,DUNE:2021hwx}. In order to lower the time and computing resources required to process the large sample used for the studies presented in this paper, a reduced version of the HD module was used for the detector simulations, the so called `1x2x6' geometry, shown in Fig.~\ref{fig:geometry}, consisting of one plane of APAs (2-APAs tall and 6-APAs long) and two cathode planes assembly (CPAs), situated one drift-length away on either side of the APAs plane in $x$. This active-volume representation, \SI{719}{cm} $\times$ \SI{1208}{cm} $\times$ \SI{1394}{cm} ($x \times y \times z$), has the same height, about half the width, and a quarter of the length of the full HD module~\cite{DUNE:2022VTX}. Wire-Cell v0\_24\_3 is used to simulate the electronics and field response, and to process the signal, along with the basic processing of the readout waveforms, which involves the deconvolution of the electric field response and noise removal in order to recover the ionization signals~\cite{wirecell_microboone}. Hit finding is performed by LArSoft by fitting Gaussian functions to peaks in the waveforms. The resulting hits are fed into Pandora pattern recognition algorithms, with the high-level reconstruction being made using LArPandora v9\_21\_12, which depends upon PandoraSDK v03-04-01~\cite{pandora_sdk}.  

With these settings, a sample of $\sim$12 million atmospheric neutrinos with energies between \SI{0.1}{GeV} and \SI{10}{GeV} was generated. The interactions are isotropically generated in the entire active volume of the reduced geometry used, with no interactions generated outside this volume. The sample contains charged current (CC) and neutral current (NC) interactions for all the neutrino flavors according to their relative cross-section ratios as computed by GENIE. An exponential energy flux spectrum ($E^{-2.5}$) with isotropic directions is used for this generation in order to have a sufficient number of interactions at higher energies. No further manipulation was done to the spectrum (\eg, reweighting with the atmospheric neutrino flux expected at the Homestake mine \cite{honda}, or considering oscillation probabilities, \etc), as the reconstruction performance results were studied separately for \nue\ and \numu; any further physics discussion is outside the scope of this paper. As such, the interaction rate differences, which can be seen in Fig.~\ref{fig:containment-repartition}, are solely shaped by the cross-section differences between different flavours, as implemented in GENIE.

MC statistical uncertainties, displayed as vertical error bars, are reported for all following figures.

\subsection{Vertex fiducialization and interaction containment}
\label{subsec:fiducialization}
A fiducialization selection was applied on the full sample, requiring the reconstructed primary vertex position to be further than \SI{20}{cm} from all the LArTPC walls. This selection: i) helps to maximize the statistical reach of the sample, while cleaning events with interactions at the edge of the active volume, which are poorly reconstructed; and ii) mimics a similar selection that will have to be used by any future physics analysis to remove neutrino interactions that are harder to distinguish from external backgrounds. Since the vertex position is reconstructed from all the hits of the event, and it is the starting point for the rest of the high-level reconstruction in the event (tracks, showers, particle identification, \etc) it was seen as a reliable reference to use in deciding on a fiducial volume (FV) selection. An additional requirement is that Pandora reconstruct at least one final state particle (track or shower), which allows for the inclusion of events that actually have detected final state particles to conduct any analysis with.

\begin{figure*}[bt]
    \centering
    \includegraphics[width=0.45\linewidth]{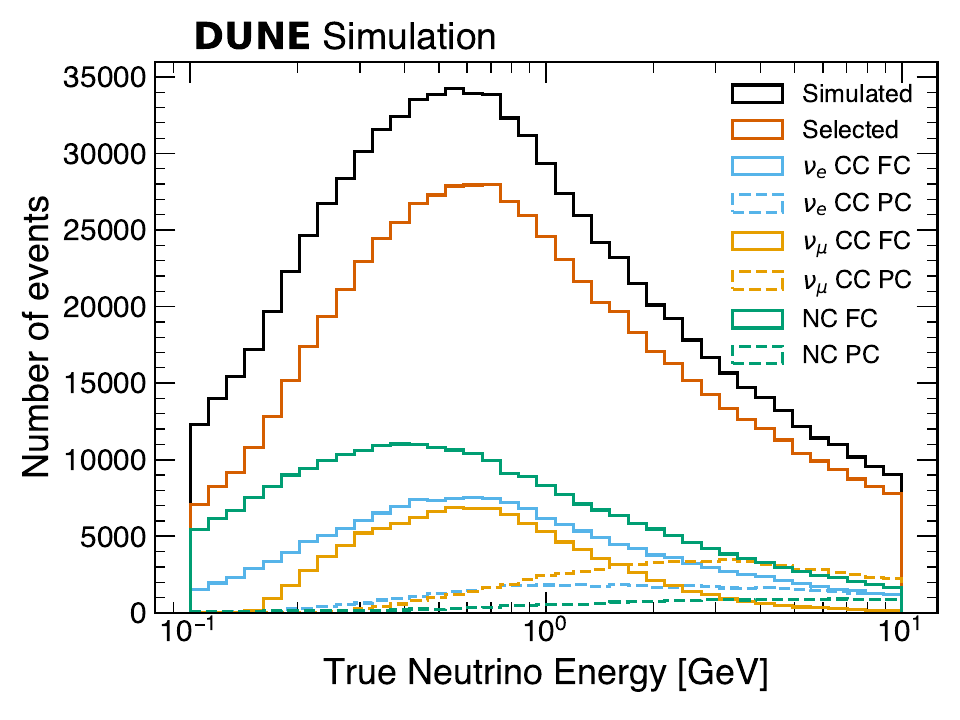}
    \includegraphics[width=0.45\linewidth]{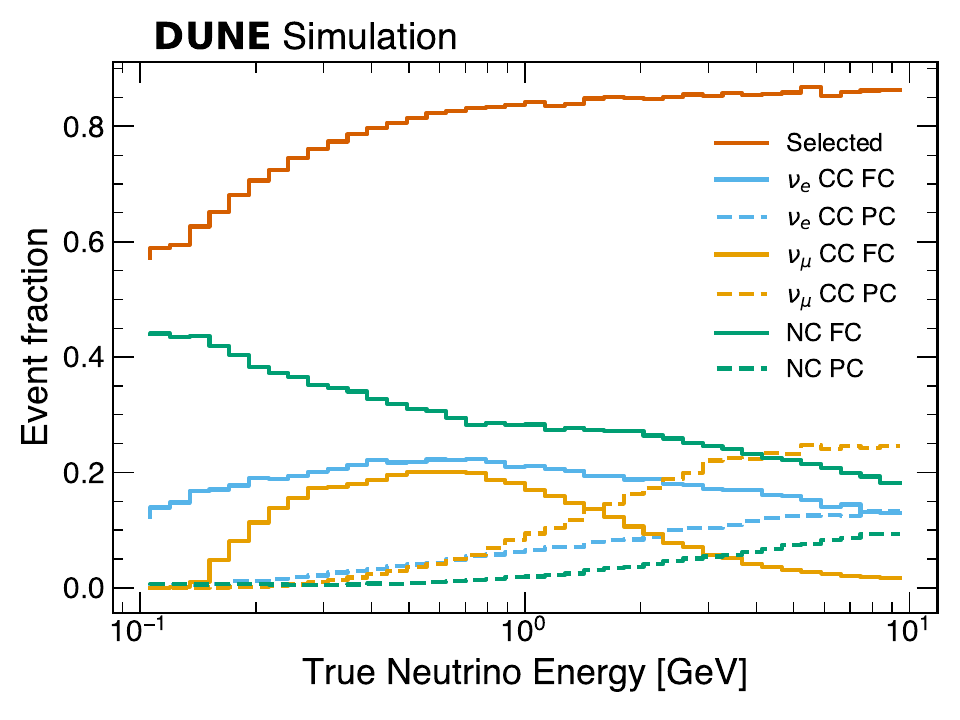}
    \caption{Neutrino energy distributions for CC and NC events separated in various categories. `Simulated' label corresponds to all simulated events, while `Selected' label refers only to events passing the fiducial volume (FV) cut as well as the reconstructed final state particle selection. The number of events (Left) and the event fraction (Right) with respect to the amount of simulated events per bin, for fully contained (FC) and partially contained (PC) events.}
    \label{fig:containment-repartition}
\end{figure*}

Atmospheric neutrino interactions can occur in the full LArTPC volume, with any incoming direction. Most of these neutrinos have a relatively low energy ($<\SI{1}{GeV}$), and consequently the full chain of interactions they produce will have a relatively small spatial extent in the detector. However, the atmospheric flux also features high energy neutrino interactions with a very large extent in the detector, for which part of the final state particles can be uncontained. So the last aspect of the selection considered in this paper is related to the containment of a neutrino interaction (\ie, whether its energy is fully or just partially deposited within the LArTPC volume). The FD-HD is expected to record both fully contained (FC) and partially contained (PC) atmospheric events in DUNE, and all of them can be considered for physics analyses. However, different reconstruction performance is expected for FC and PC events, in particular for the energy resolution. In order to best capture the effects of containment on atmospheric neutrino physics in DUNE, the analysis sample was split into FC and PC sub-samples, based on the following chosen criteria: the FC (PC) sub-sample contains neutrino events for which no (at least one) hit belonging to a reconstructed final state particle by Pandora is present outside the fiducial volume. It should be noted that no radiological background (arising from the natural contamination of radioactive isotopes in detector materials and the surrounding cavern) was considered in this study, as they are primarily visible in neutrino LArTPCs via MeV-scale signatures~\cite{Caratelli:2022llt}, which are outside the energy range studied in this paper. 

Figure~\ref{fig:containment-repartition} shows the effect of all these selections as a function of the true neutrino energy. The FV and the `at least one reconstructed final state particle' selections filter mostly low-energy neutrinos. The splitting into the FC and PC categories has, as expected, a very strong energy dependence, since most of FC events correspond to low-energy neutrinos (below \SI{1}{GeV}). In addition, there is also a neutrino type dependence, because $\nu_\mu$ events are less easily contained than $\nu_e$, as the rate of energy loss of the muon in the liquid argon is much smaller than that of the electron (\eg, a \SI{1}{GeV} muon travels for $\sim$34 radiation lengths in LAr).


%% file: AtmReco.tex
\section{Atmospheric neutrino event reconstruction}
\label{sec:atmReco}

\input{vtx_reco}
\input{part_reco}
\input{nu_eReco}
\input{nu_dirReco}
\input{nu_flavReco}

%% file: vtx_reco.tex
\subsection{Neutrino-interaction vertex finding}
\label{subsec:vtx}

The neutrino interaction vertex is the point from which all charge depositions in the detector emanate, and many Pandora algorithms (as well as other high-level reconstruction observables like PID, direction, \etc) leverage it as an anchor point to reconstruct the emitted final-state particles. 
The original~\cite{pandora_sdk} Pandora vertex-location finding algorithm was re-written in DUNE to use a convolutional neural network (CNN)~\cite{DUNE:2022VTX}. The architecture of the network is kept exactly the same for beam and atmospheric neutrino events, with two stages of processing: pass one is a coarse search to define the correct region within an event for the vertex position, pass two is a more detailed search within that region. The only difference between beam and atmospheric vertex finding is that the network is trained separately on the two samples, for two reasons. First, a CNN trained only on beam events leads it to give a specific importance to the beam direction with respect to any other direction, leveraging this information of the beam direction to easily break ambiguities in the vertex location for these type of events. However, as atmospheric neutrinos can come from any direction, using a CNN trained on beam events would introduce a bias. Second, the atmospheric neutrinos have a different flux than beam neutrinos, with a greater fraction of neutrinos expected below \SI{1}{GeV}. 

\begin{figure*}[tb]
    \centering
    \includegraphics[width=\linewidth]{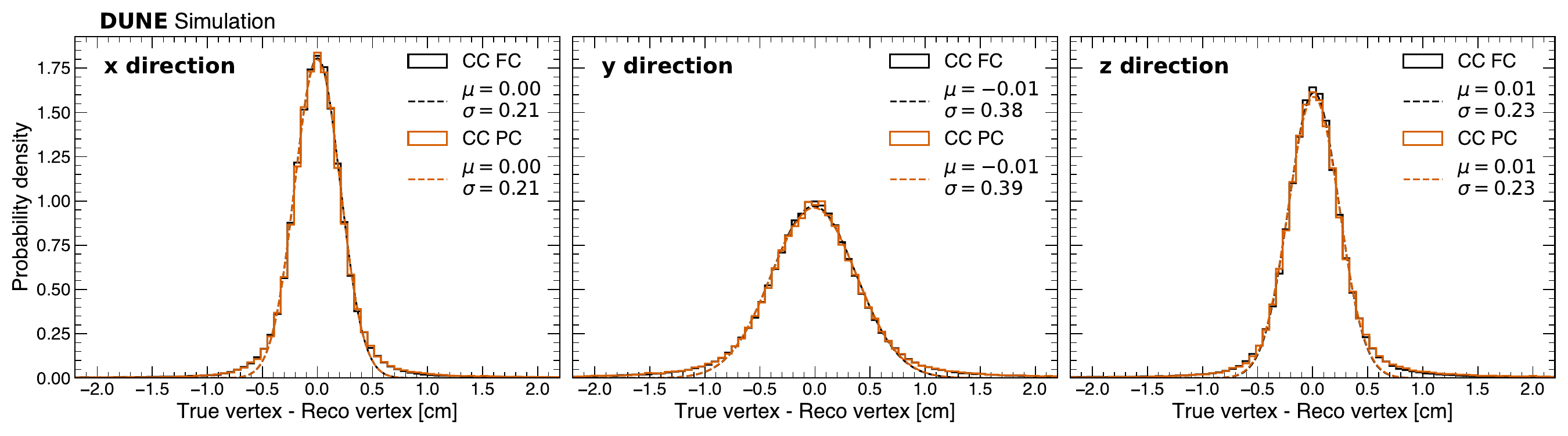}  
    \caption{Resolution distributions of the reconstructed primary interaction vertex, for (from Left to Right) the $x$, $y$, and $z$ directions, for CC interactions, with all flavors combined. The mean ($\mu$) and standard deviation ($\sigma$) for each distribution are also listed.}
    \label{fig:vtx-res}
\end{figure*}

The CNN training for the atmospheric events was performed directly on a subset of \SI{600}{K} events (from the whole sample), reconstructed up to the stage of the 2D hit reconstruction, without any further selection. The training was performed in a staged way (as for the beam~\cite{DUNE:2022VTX}), one for each of the two passes of the vertex reconstruction algorithm: the first pass was retrained first before being used to retrain the second pass. The obtained vertex resolution is shown in Fig.~\ref{fig:vtx-res}, for CC only (as they are guaranteed to have vertex activity), FC and PC events, in each direction. The performance is comparable to that obtained for beam events~\cite{DUNE:2022VTX}, with resolutions below  \SI{1}{cm} for each plane. The degraded performance in the $y$ axis case can be explained by the orientation of the collection plane's wires in particular: they are collinear with $y$, and thus provide less spatial resolution in this direction. It can be seen that the containment of the interactions has a minimal impact on this performance, and the same was found for the flavor of the neutrino. 

\begin{figure*}[tb]
    \centering
    \includegraphics[width=\linewidth]{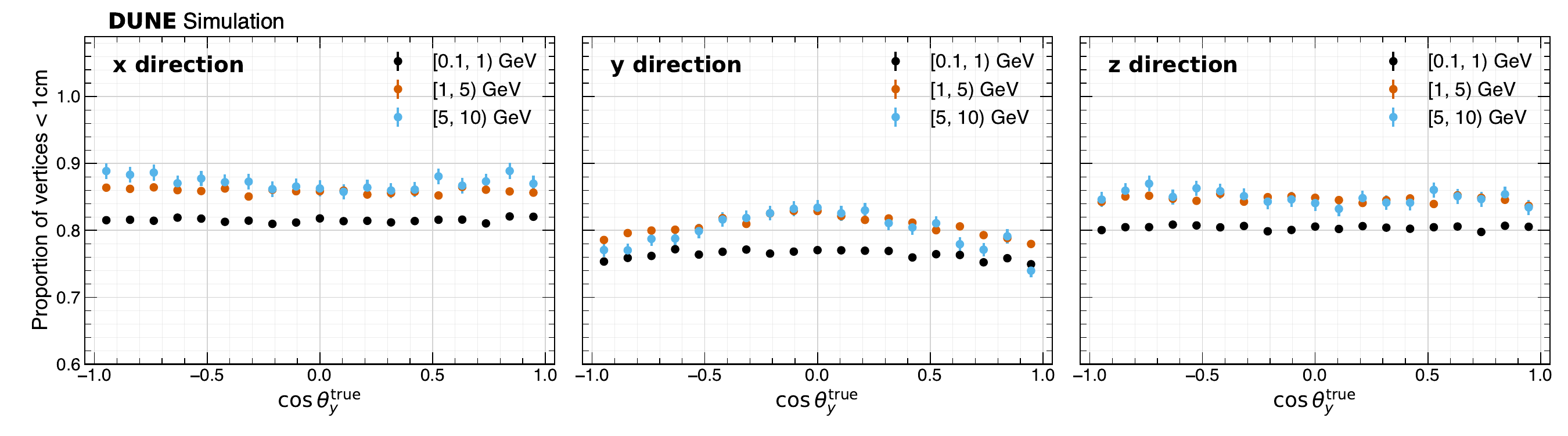}
    \caption{Proportion of vertices reconstructed within \SI{1}{cm} in a given direction from the true vertex, as a function of the zenith angle \tetay{}, for (from Left to Right) the $x$, $y$, and $z$ directions. Results are shown for three neutrino-energy intervals, for CC FC events only.}
    \label{fig:vtx-res-y-all}
\end{figure*}

Figure~\ref{fig:vtx-res-y-all} shows the proportion of events for which the vertex is reconstructed within \SI{1}{cm} of the true vertex location in a given direction as a function of the true neutrino zenith direction, in three neutrino-energy intervals: 0.1 -- 1, 1 -- 5, 5 -- 10~\GeV. This efficiency allows the characterization of the quality of the vertex reconstruction for a given subset of events. The \SI{1}{cm} threshold is chosen as this is in the large majority of cases an accuracy sufficient to ensure the quality of downstream Pandora reconstruction. An increased vertex reconstruction efficiency is observed for neutrinos propagating orthogonally to the $y$ direction, with average efficiencies of 82\% and 80\% \vs\ 76\% (87\% and 85\% \vs\ 80\%) for the lowest (highest) energy bin shown in $x, z$ and $y$ directions, respectively. The dependency on neutrino propagation direction increases with the neutrino energy, as expected by the fact that the produced final state particles will be more and more boosted in the true neutrino direction as the energy increases, and thus show more clearly the effects of the neutrino directionality on the reconstruction performance. 


%% file: part_reco.tex
\subsection{Particle reconstruction}
\label{subsec:partreco}


\subsubsection{Efficiency, purity, and completeness}
\label{subsec:atmbeam}

The metrics used for assessing the particle reconstruction performance in this subsection are the following: 

\begin{itemize}
    \item \textbf{Purity:} the fraction of hits in the reconstructed particle that are matched with that of the corresponding generated particle, $N_\mathrm{recoHits}^\mathrm{matched}/ N_\mathrm{recoHits}$;
    \item \textbf{Completeness:} the fraction of hits in the generated particle that are matched with the corresponding reconstructed particle, $N_\mathrm{genHits}^\mathrm{matched}/ N_\mathrm{genHits}$;
    \item \textbf{Efficiency:} the fraction of a type of target MC generated particles matched to a reconstructed particle. In order to ensure that the reconstructed PFP is associated with a single MC particle, the matching criteria used are: i) at least 5 shared hits, in order to avoid the metric being skewed by small clusters from things like neutron capture; ii) and purity and completeness levels of at least 50\%, to ensure that the majority of the hits in the PFP are associated with the MC particle, while excluding matches between target MC particles and small, fragmented reconstructed particles. Double counting is also avoided in the way the counting is set.
\end{itemize}
A subsample of $\sim$\SI{1.5}{M} events from the large \SI{12}{M} sample has been used for the studies in this section. The events are required to pass the vertex fiducialization selection, without any further requirement on containment.

\begin{figure*}[tb]
	\centering
    \includegraphics[width=0.3\linewidth]{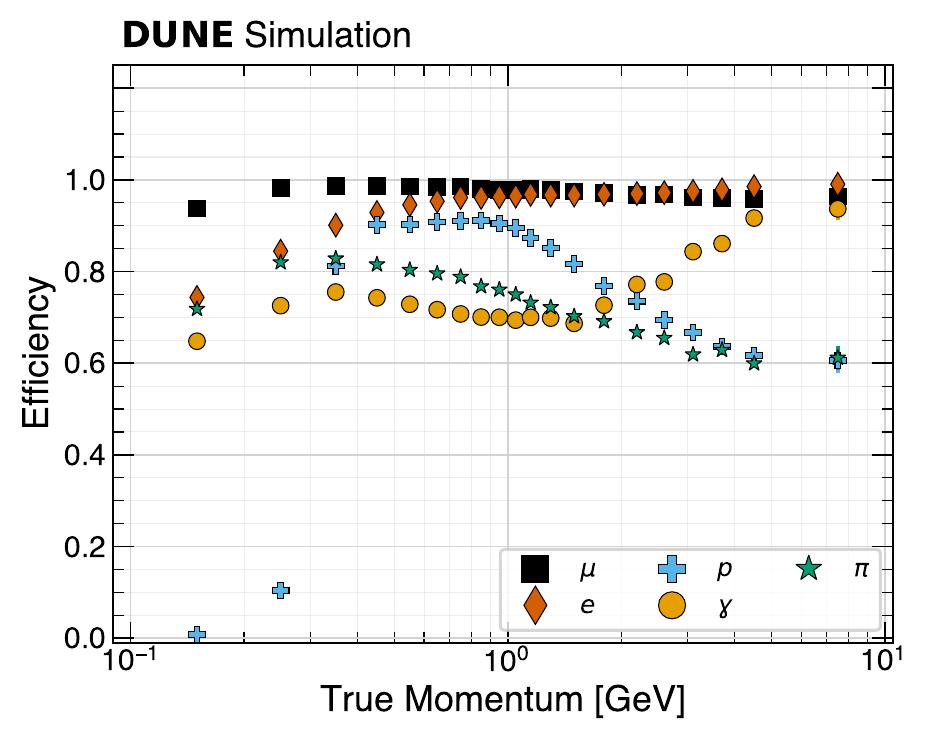}
    \includegraphics[width=0.3\linewidth]{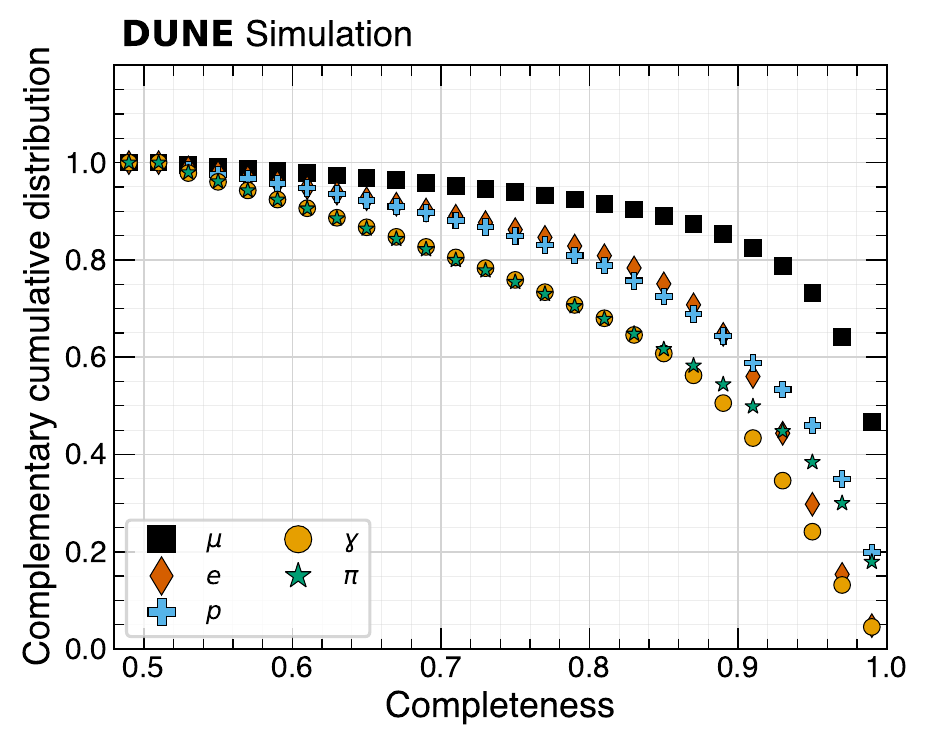}
    \includegraphics[width=0.3\linewidth]{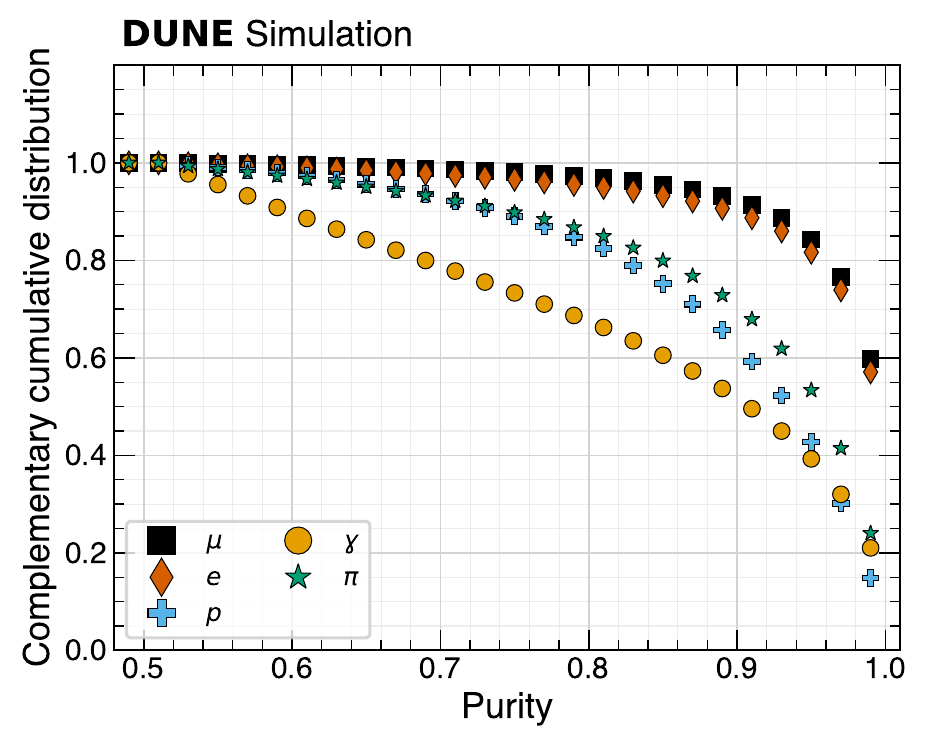}
	\caption{Efficiency versus true particle momentum (Left), and complementary cumulative distributions for completeness (Middle) and purity (Right). The complementary cumulative distributions represent the fraction of events with completeness/purity above a certain x-value.}
	\label{fig:part_reco}
\end{figure*}

The results are shown in Fig.~\ref{fig:part_reco}, combined for CC and NC neutrino interactions. The reconstruction efficiency of muons is the highest among all particles over the full energy range shown, and rather flat. For protons, the efficiency shows a sharp decrease above $\sim$\SI{1}{GeV}, caused by the fact that protons below this energy mostly leave tracks in the detector, while above they tend to produce hadronic showers, which are more difficult to reconstruct. Additionally, high-energy protons are associated with complex deep inelastic scattering (DIS) topologies in which most of the hits are clustered together into a large shower by Pandora. 

At low energies, below $\sim 300\,$MeV, the efficiency drops for all particles: for muons and electrons, this is due to high-energy neutrino interactions where the charged lepton carries a smaller fraction of the energy and may be clustered with more energetic hadronic or photonic showers. For protons and photons, the same issue occurs, compounded by the intrinsic difficulty in detecting particles that leave very few hits and are easily absorbed into nearby lepton PFPs. At high energies, above $\sim 2\,$GeV, photon reconstruction efficiency improves, as they tend to carry a significant fraction of the neutrino energy and are more easily matched to reconstructed objects due to the matching algorithm prioritizing shared hits. Pions exhibit behavior similar to protons, though slightly worse at low energy, as their reconstruction is more strongly impacted by the complexity of the interaction topology they are usually associated with. Muons remain the best-reconstructed species across all metrics due to their long, clean track-like signatures. All these considerations are reflected also in the completeness and purity results, with highest performance for muons and poorest for photons.

\subsubsection{Particle identification method}
\label{subsubsec:pid}

The 3D high-resolution tracking and the 4$\pi$ calorimetry capabilities of the LArTPCs are what distinguishes them from other neutrino detection technologies. The particle identification that these characteristics will enable in DUNE should allow for a superior neutrino direction and energy reconstruction performance. The measurement accuracy of these quantities improves when knowing precisely the type of the particle (muon, pion, electron, proton, \ldots) depositing energy in the LAr, as different particles have different energy deposition profiles, and therefore different amounts of charge recombination (\ie, ionization electrons that recombine with argon ions). Identifying the particles and thereby correcting/calibrating their kinetic energy (\Ekin) and extracting their momenta accordingly improves neutrino reconstruction, as shown in Secs.~\ref{subsec:nudirection} and \ref{subsec:nuenergy}. Two phases can be distinguished in the particle identification (PID) process: the separation between track-like and shower-like objects, and then the identification among these of the different particle types. 

\begin{figure*}[tb]
    \centering
    \includegraphics[width=0.45\linewidth]{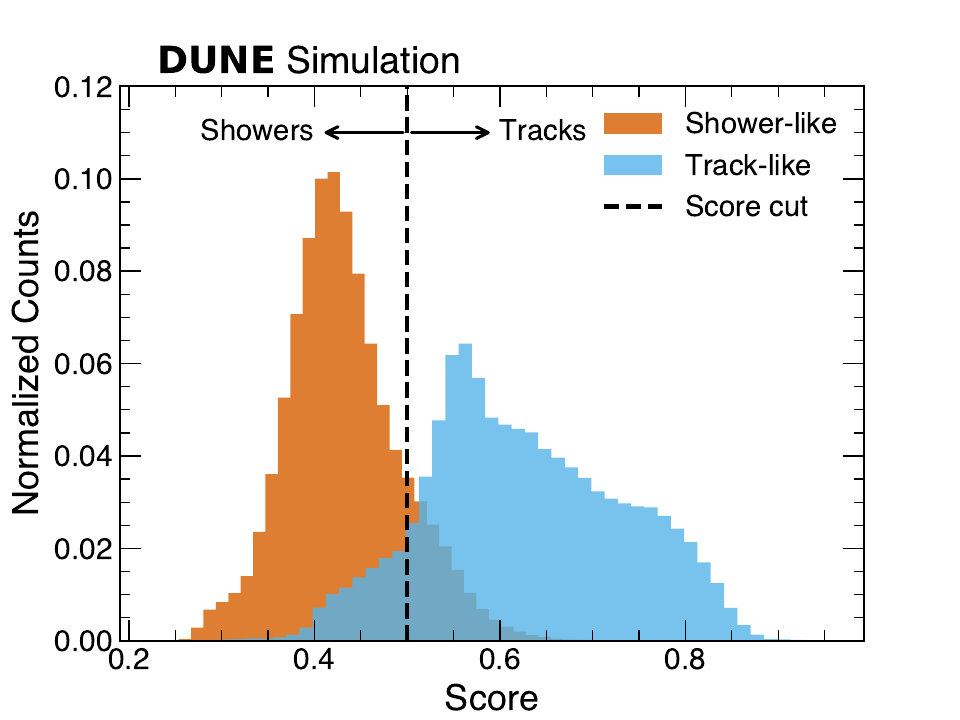}
    \includegraphics[width=0.45\linewidth]{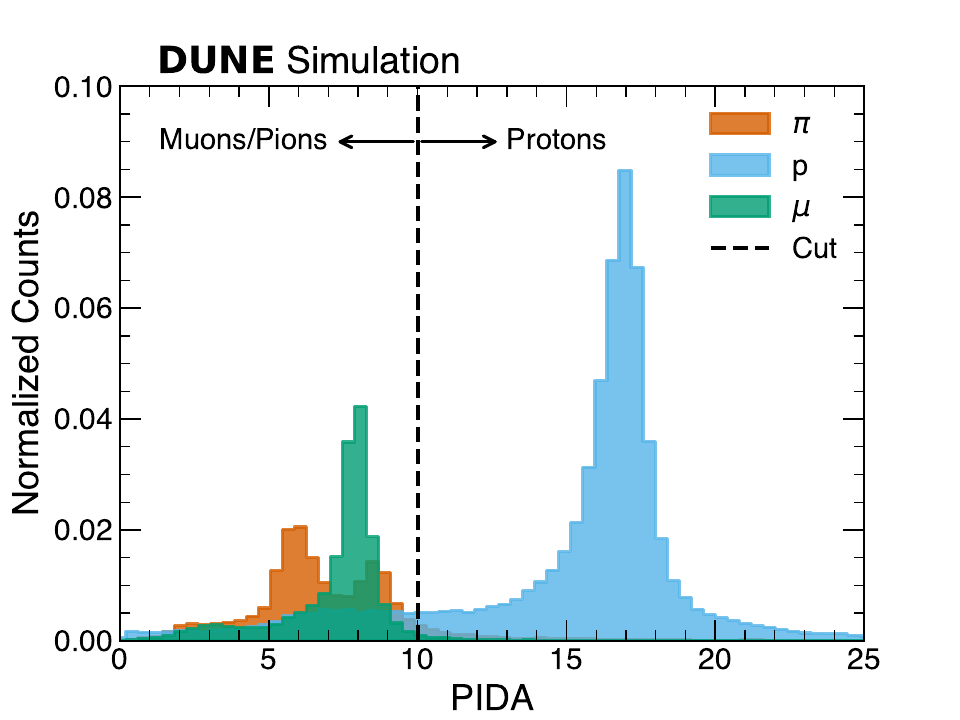}
    \caption{Distributions of the classification scores: (Left) Track-Shower BDT scores; (Right) PIDA scores for Pandora tracks split into the true particles they are associated with.}
    \label{fig:classification}

\end{figure*}

The separation between track-like and shower-like objects is performed by a Boosted Decision Tree (BDT) integrated within Pandora~\cite{pandora_uboone} that was retrained on a subset of the atmospheric sample. Hyper-parameters such as the number of training examples and the learning rate were optimized based on a subsample of $\sim\SI{1.5}{M}$ events. The final training was made on a distinct sample containing $\sim\SI{190000}{PFPs}$ reconstructed with purity and completeness above 0.5. The obtained results can be seen in the left panel of Fig.~\ref{fig:classification}, following the convention established by MicroBooNE~\cite{uboone_discrim} for the definition of the Track-like and Shower-like particles. Applying a cut on the BDT score at 0, allows to select tracks with an efficiency of \SI{89}{\percent} while rejecting \SI{91}{\percent} of the showers.

The approach taken for performing PID, to separate highly ionizing particles (e.g. protons) from minimum ionizing particles (e.g. muons), was to deploy the PIDA~\cite{PIDA} phenomenological model to characterize the Bragg peak. A PIDA score was calculated, which relies on the power-law dependence of \dedx on the particle residual range, as it reaches the end of its travel within a medium:
\begin{equation}
    \frac{dE}{dx} \simeq AR^b
\end{equation}
where $R$ is the residual range, $b$ is a constant varying weakly with the particle type (-0.42), and $A$ is a proportionality coefficient, which varies with the particle type, and is to be measured. The PIDA score is calculated for each reconstructed Pandora track on its last \SI{30}{cm} (or less if the track is shorter) as:
\begin{equation}
    \text{PIDA} = \frac{1}{N_\text{points}}\sum_i^{N_\text{points}}\left(\frac{dE}{dx}\right)_{\text{reco}, i}R_i^{0.42}.
\end{equation}

The results displayed in the right panel of Fig.~\ref{fig:classification} show a good separation between the proton and the muon/pion distributions. A selection requirement of $\text{PIDA}>10$ retains $\sim$87\% of the protons with a purity of 91\%. However, it can also be seen that there is very little separation power between the charged pions and muon tracks, which can be understood from them having a very similar \dedx profile. The specific double peak feature for pions was understood by the two types of pions tracks reconstructed in the detector. One, corresponding to stopping pions, will have the expected energy deposition profile from an ionizing particle of this mass, and as a consequence, the associated PIDA score will be computed correctly. The other, from pions undergoing inelastic hadronic interactions, have the end of their track not corresponding to a particle at rest, and therefore, the residual range $R$ used for the PIDA calculations will be wrong, and this will lead to a shift to lower PIDA values of the particle.
Even though the possibility to explicitly tag pions undergoing an inelastic hadronic interaction with PIDA and apply a specific treatment to them in the reconstruction seems promising, it has not been applied here and it is left for future work. Similarly, an accurate discrimination between charged pions and muons should be possible by considering additional geometric features such as the track scattering or the track lengths, but do not consider it in the following studies. 

\subsubsection{Direction reconstruction}
An accurate determination of individual final-state particle direction is essential for the reconstruction of atmospheric neutrinos. This is because for these types of neutrinos, there is no knowledge of the traveling direction, hence no knowledge about the distance traveled, which directly impacts any oscillation measurement. This reconstruction of the direction is made by two separate algorithms, one for tracks, and one for the hadronic and electromagnetic showers.
In the track reconstruction module, a Principal Component Analysis (PCA) is first applied to define a local 3D coordinate system aligned with the track's principal axis. Subsequently, a sliding-window linear fit is performed on the constituent 2D hits. This method reconstructs the track using a series of local segments, which are combined to yield 3D trajectory points and estimate the overall track direction. In contrast, the shower module reconstructs direction by applying PCA directly to the spatial distribution of the 3D hits.
The actual pointing of the shower is made in the direction of the largest spatial extent.

\begin{figure*}[tb]
    \centering
    \includegraphics[width=.9\linewidth]{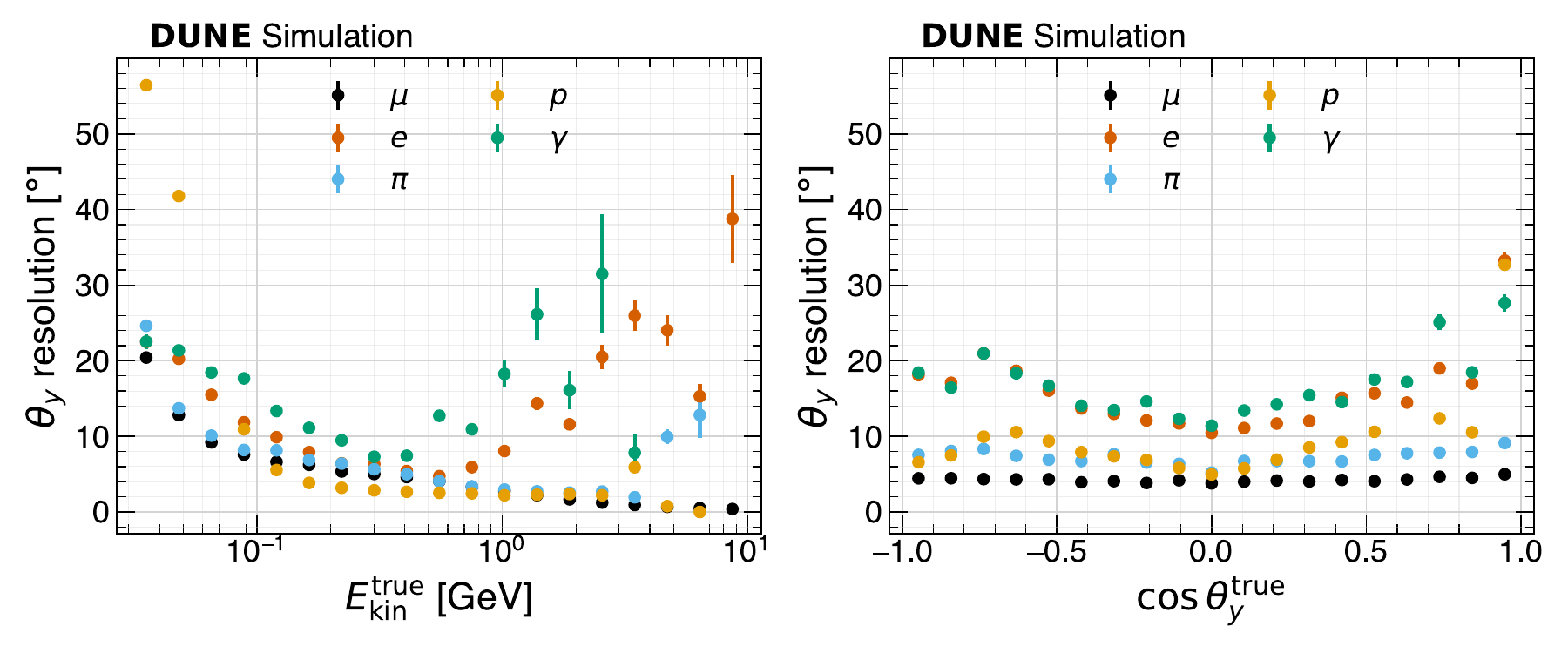}
    \caption{Resolution of reconstructed particle's angle with the $y$ direction ($\theta_y$), as a function of (Left) particle's true kinetic energy \Ekintrue, and (Right) of $\cos(\theta^{\rm{true}}_y)$, for muons, pions, electrons, protons, and photons.}
    \label{fig:track-angle-res}
\end{figure*}

The left panel of Fig.~\ref{fig:track-angle-res} shows the resolution on the reconstruction of particles direction with respect to the $y$ direction (most relevant for any atmospheric neutrino oscillation analysis) for all types of particles. Both the fully and partially contained particles are considered as the containment was checked to only have a minimal impact in this case. While the best direction resolution is obtained for muons, which have a resolution below \SI{5}{\degree} at a kinetic energy of \SI{1}{GeV}, the following general trends can be observed:
\begin{itemize}
    \item For all particles, the resolution is relatively poor at low-\Ekin{} (below \SI{100}{MeV}), because the tracks or showers are shorter and the accurate reconstruction of their direction is more difficult. This is especially the case for proton tracks, which only span very few centimeters at these energies.
    \item For shower-like particles, the best resolution is obtained for \Ekin{} between \SI{100}{MeV} and \SI{1}{GeV}, where the individual track/shower reconstruction is easy as they are all long enough, and the events are not very crowded.
    \item A degradation of resolution is observed at higher \Ekin{} for shower-like particles, as the events become more and more crowded, and the accuracy in the reconstruction of the individual particles decreases. This is mainly due to the over-clustering of the particle hits together in the last parts of the reconstruction process.
\end{itemize}
In general, electrons and photons show lower performance compared to other particles, especially at higher energies, because their showers are more likely to be confused with the overall hadronic system while tracks can more easily be separated from hadronic showers.

The right panel of Fig.~\ref{fig:track-angle-res}, in which a selection cut \Ekinreco$>$\SI{10}{MeV} was applied to all particles, shows that there is also a dependency of the angular resolution on the true initial direction of the particle itself. Muons direction resolution has little dependence on the true muon direction. On the contrary, other particles show a slightly stronger dependency, with the best resolutions obtained for particles orthogonal to the $y$ axis. A loss of resolution is observed for tracks very aligned with $y$ (or -$y$) direction, because of the lack of hits on the W view, whose wires are themselves oriented along the $y$ direction, thus not providing any information for these tracks. 



\subsubsection{Energy reconstruction: muons}
\label{subsubsec:erec-muons}
Reconstructing the muon energy precisely is of critical importance in order to accurately reconstruct the energy of muon neutrinos.
Two methods are used for the energy reconstruction of muons in DUNE. If the reconstructed track is fully contained, the kinetic energy \Ekin\ is calculated by range using the Constant Slow Down Approximation~(\csda{})~\cite{pdg_csda}. However, when the muon track is uncontained, it is not possible to rely on such a method as it can only be applied to stopping muons. Escaping muons are not rare in the case of atmospheric neutrino interactions given the energy flux, the geometry of the detector and the direction of the neutrinos. This is why, in the context of the atmospheric reconstruction, additional muon energy reconstruction methods relying on the \mcsname~(\mcs) were intensively studied. One method, \erecochi, is based on the OPERA and ICARUS~\cite{OPERA_chi2,ICARUS_chi2} experiments implementation, while another one, \erecollhd, is based on that of the MicroBooNE~\cite{uboone_llhd} experiment. While both methods are implemented in the DUNE reconstruction software at present, in the context of this paper, the focus is on the results obtained with the \erecollhd\ method, which was found to perform consistently better (\ie, smaller biases and improved resolutions) than the \erecochi\ one in the case of atmospheric neutrinos.

In general in \mcs\ approaches, the track is first divided into segments of length $x$, followed by a straight-line fit to the space points in each segment. The scattering angle between each pair of connected segments is calculated, and the obtained distribution of scattering angles for all segments belonging to a track is fitted with a Gaussian centered at zero, from which the standard deviation is computed. The procedure is repeated for six different segment lengths between \SI{10}{cm} and \SI{60}{cm}~\cite{OPERA_chi2}. In the end, the scattered angle of the muon track, $\theta_0$, is the standard deviation of the Gaussian distribution and given by the Highland formula:
\begin{equation}
\label{eq:highland}
    \theta_0 = \frac{13.6\MeV}{\beta cp}z\sqrt{\frac{x}{X_0}}\left[1+0.038\ln{\left(\frac{xz^2}{X_0\beta^2}\right)}\right],
\end{equation}
where $\beta c$, $p$, $z$ are the particle's velocity,  momentum, and charge number, respectively, and $X_0$ is the radiation length of LAr (assumed constant and equal to \SI{14}{cm}). For the \erecollhd\ method in particular, only the scattered angles $\Delta\theta$ of \SI{10}{cm} segments are considered. Given the normal probability of the scattering angle
\begin{equation}
    f(\Delta\theta) = \frac{1}{\sqrt{2\pi \theta^2_0}} \exp{\left[-\frac{1}{2}\left(\frac{\Delta\theta}{\theta_0}\right)^2\right]},
\end{equation}
the likelihood is estimated by the product of $f(\Delta\theta_i)$ over all $n=n_\text{segments}-1$ scattering angles $\Delta\theta_i$~\cite{uboone_llhd}, and the momentum is measured by minimizing the negative log likelihood:
\begin{equation}
\label{eq:enllhd}
    -\ln(L) = \frac{n}{2} \ln(2\pi) + \sum_{i=1}^n \ln(\theta_{0,i}) + \frac{1}{2}\sum_{i=1}^{n}\left(\frac{\Delta\theta_{i}}{\theta_{0,i}}\right)^2.
\end{equation}

In order to compute the expected scattering angle using Eq.~\ref{eq:highland}, Eq.~\ref{eq:enllhd} takes into account the momentum of the particle in each segment. Therefore, the energy loss throughout the segments needs to be taken into account. In the minimization process, this energy loss is computed using the most probable energy loss $dE/dx$ value (MPV) in liquid argon~\cite{pdg_csda}. In addition, Ref.~\cite{uboone_llhd} has tuned Eq.~\ref{eq:highland} for LAr, and found that the value of \SI{13.6}{MeV} depends on the particle's momentum. This tuning, however, depends on how the scattering angle is being computed and how the energy loss is being taken into account. Because of this, and following a similar approach as Ref.~\cite{uboone_llhd}, Eq.~\ref{eq:highland} was modified to the form
\begin{equation}
\label{eq:highland_tunned}
    \theta_0 = \frac{\kappa(p)}{\beta cp}z\sqrt{\frac{x}{X_0}}\left[1+0.038\ln{\left(\frac{xz^2}{X_0\beta^2}\right)}\right],
\end{equation}
in which the $\kappa(p)$ parameter is fitted as function of the true momentum, in each segment.

\begin{figure*}[tb]
    \centering
    \includegraphics[width=0.45\linewidth]{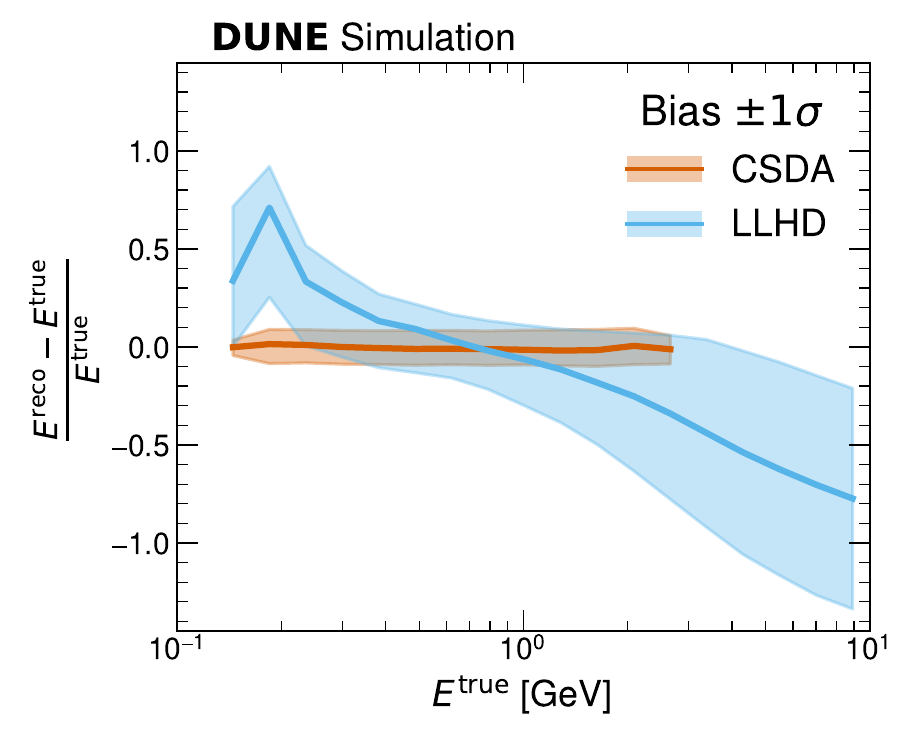}
    \includegraphics[width=0.45\linewidth]{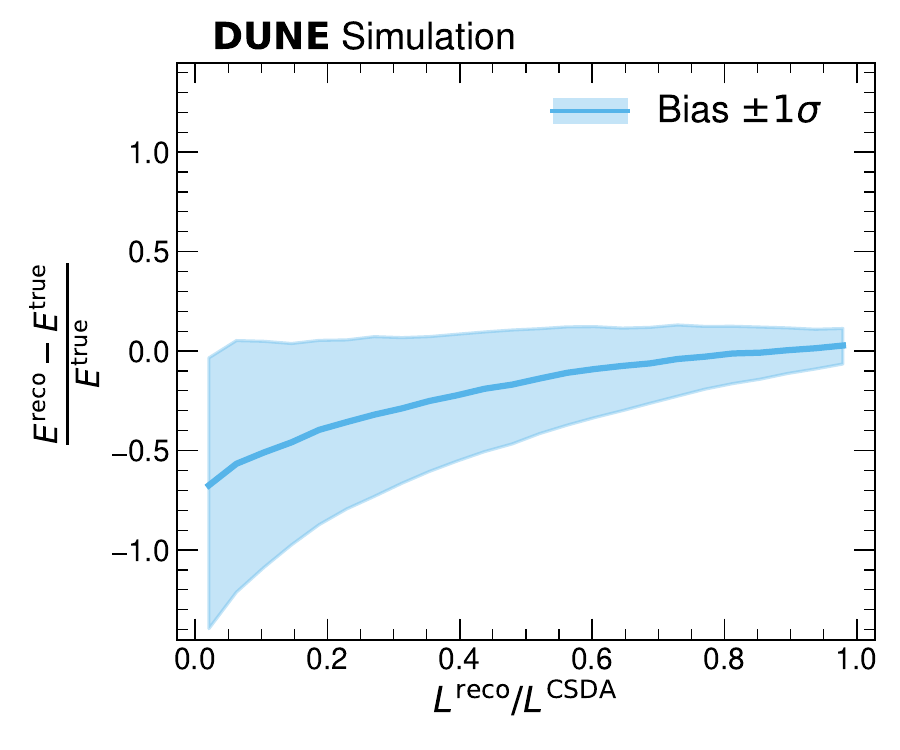}
    \caption{Energy resolution and bias: (Left) versus true muon energy for the \csda\ and \erecollhd\ methods, when requiring the longest track to be fully or partially contained, respectively; (Right) versus relative length inside the TPC for the \erecollhd\ method and partially contained longest tracks only. The $L^\text{\csda}$ refers to the expected muon track length computed using the \csda\ method on the true muon momentum.}
    \label{fig:ereco_res_bias}
\end{figure*}

Figure~\ref{fig:ereco_res_bias} shows the resolution and bias distribution of reconstructed energy when using the \csda\ and \erecollhd\ methods in its left panel. The CSDA method shows extremely good performance ($\sim\SI{2}{\percent}$ resolution) across the energy range. A cut is applied at \SI{3}{GeV} because too few muons are contained in the simulated geometry above this energy. With the MCS LLHD, the resolution is of course degraded, but this method allows to obtain a \SI{25}{\percent} energy resolution on the majority of uncontained muons below \SI{1}{GeV}, which allows to obtain a reliable energy measurement even for partially contained \numu\ CC events. At higher energies, this method becomes limited by the intrinsic resolution of the detector that becomes unable to measure the small scattering angles of the muon, hence leading to an underestimation of the muon momentum and a large negative bias.

The simulation used for this paper uses only about $1/8$th of the full detector geometry volume (with full vertical length). In the full-size detector, muon tracks will be longer, and two things are expected to change: (1) more tracks will be tagged as contained and (2) for exiting tracks, for a fixed energy, the fraction of the track-length inside the TPC will increase. Therefore, a better bias and resolution are expected when using \csda\ in case (1) or \mcs\ in case (2). This can already be seen in the right panel of Fig.~\ref{fig:ereco_res_bias}, which shows the resolution and bias of the \mcs\ calculation versus the relative length inside the TPC, defined by the ratio of the reconstructed length to the expected length computed using the \csda\ method on the true muon momentum. One can notice the improvement in energy reconstruction when the tracks have a higher relative length inside the TPC. 

In Fig.~\ref{fig:ereco_res_bias}, and all other figures of the same type in the rest of the paper, the relative resolution and bias are shown as a band plot where the main line corresponds to the relative bias while the $\pm\si{1}{\sigma}$ asymetric resolutions are shown as the distance between the bias line and the upper (lower) band extremity respectively. They have been computed in the following way: 

\begin{itemize}
    \item \textbf{Relative bias}: is calculated as the mean value of the $(\text{reco} - \text{true})/\text{true}$ ratio, for a given variable, for all events falling into a given bin.
    \item \textbf{Relative resolution:} is estimated following several steps: i) the events are sorted into bins of a given variable; ii) for all events within each bin, the $(\text{reco}/\overline{\text{reco}})$ distribution is evaluated, where $\overline{\text{reco}}$ is the mean value of the $\text{reco}$ variable within the bin; iii) for each bin, the \SI{16}{\percent} and \SI{84}{\percent} quantiles of the distribution are evaluated; iv) this inter-quantile measurement is divided by 2, corresponding to the definition of one standard deviation in the case of a gaussian distribution.
\end{itemize}



\subsubsection{Energy reconstruction: electrons}
\label{subsubsec:erec_others}
Electrons with an energy in the \si{GeV} range produce electromagnetic showers in the detector that develop as the electron emits bremsstrahlung gamma rays, which in turn create electron/positron pairs. The typical extent between the vertex and the maximum of energy deposition of the shower is given by
\begin{equation}
    l \simeq X_0\log{(E/E_c)},
\end{equation}
where $X_0 = \SI{14}{cm}$ is the radiation length and $E_c=\SI{32}{MeV}$ is the critical energy for electrons. It is calculated that for an electron energy $E=\SI{3}{GeV}$, this length will be around \SI{68}{cm}. However, photons can be radiated quite far from the main shower, and there are a lot of event-to-event variations in the actual shower profile. It has in fact been shown that for \SI{1}{GeV} (\SI{3}{GeV}) electrons, on average \SI{97}{\percent} of energy is deposited within a distance of \SI{1.5}{m} (\SI{1.7}{m}) from the primary vertex~\cite{lartpc-perfs}. This indicates that most of the electromagnetic showers produced by electrons will be contained in the detector. As such, it is possible to reconstruct the energy of these electrons by using a calorimetric measurement of the total energy deposited in the shower. 

\begin{figure*}[tb]
    \centering
    \includegraphics[width=\linewidth]{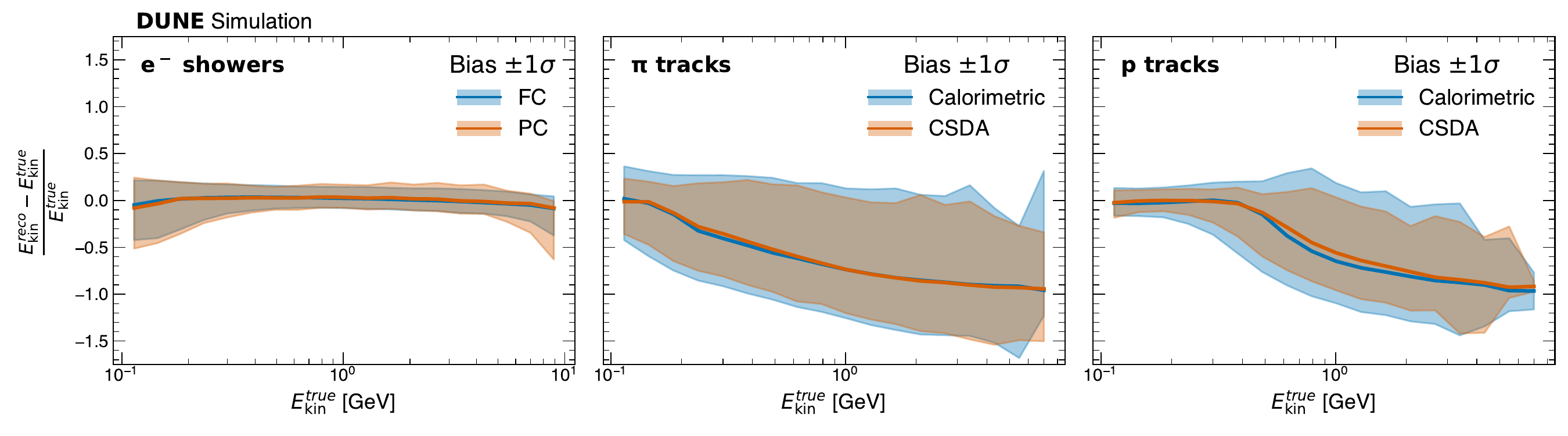}
    \caption{Energy resolution and bias as a function of the true energy for: (Left) electron showers in fully (FC) and partially (PC) contained neutrino events; and for pion (Middle) and proton (Right) tracks in FC events only, when using using two energy reconstruction methods.}
    \label{fig:resolution_all}
\end{figure*}

The energy reconstruction of showers is performed the following way. For each of the three LArTPC views, a calorimetric sum of all the hits constituting a given PFP is calculated. This calorimetric sum corrects for the charge attenuation during the drift and an average recombination factor. A charge to energy conversion is also applied according to some fixed gain calibration factors. Among the three readout planes of the DUNE HD detector, the $w$ collection plane is the one with the highest signal-over-noise ratio and as a consequence the most accurate in estimating the energy. It even provides a more accurate energy estimate than the particle-by-particle "best view" (\ie, the view that collected the highest number of hits from a given particle). As a consequence, it is the W view that was used to perform the calorimetric estimation of the energy for the results shown here. The obtained resolution for electron showers is shown in the left panel of Fig.~\ref{fig:resolution_all} for both FC and PC events. The PC events yield very similar results to the FC one, at the exception of some broadening in the energy resolution at higher energies. At low electron energies, the energy resolution is limited by the detection thresholds of the particles in the shower as several soft electrons produced in the shower do not deposit enough energy to be correctly detected as hits. In the current reconstruction, the hit detection threshold corresponds to roughly \SI{300}{keV}, but even with hit-finding thresholds as low as \SI{100}{keV} part of the deposited energy in shower could still not be detected ~\cite{lartpc-perfs}. The best performance is obtained for \SI{1}{GeV} electrons, with an energy resolution around \SI{15}{\percent}. An energy bias is present at energies $<\SI{200}{MeV}$ due to the detection inefficiencies before reaching 0 for all the higher energies.

\subsubsection{Energy reconstruction: charged hadrons}
Both protons and charged pions deposit energy continuously in the detector through ionization, leaving tracks in the detector just like muons. But there are some differences to take into account with respect to the muon case.

First, protons are often non-relativistic in DUNE due to their large mass. As a consequence, their rate of energy deposition is larger by a factor of six with respect to muons, which leads to much shorter tracks. Moreover, this higher rate of energy deposition implies a higher charge-recombination rate. Second, charged hadrons can undergo hadronic (inelastic) scatterings in LAr, which become relevant for pions and protons with energy larger than a few hundred \si{MeV}. The typical exchange of energy in this kind of process is relatively large, with a cross-section that is relatively particle and energy independent and with  an associated interaction length in DUNE around \SI{1}{m}~\cite{lartpc-perfs}. The scatterings cause loss of visible energy via nuclear breakup or recoil, or by transfer of energy to neutral or very massive charged particles, which are left undetected in the detector. Consequently, if using a calorimetric energy measurement (in which the energy is calculated as the sum of the reconstructed charge of each hit of the particle's track, corrected for recombination and drift attenuation effects), the obtained energy resolution for these charged hadrons is bimodal: there is one peak around \Ekinreco/\Ekintrue $\sim$1 corresponding to particles not undergoing hadronic interaction, and  a second wider distribution below \Ekinreco/\Ekintrue $\sim$0.5, corresponding to particles undergoing hadronic interaction which leads to a loss in visible energy. 

The \csda\ method suffers from the same limitations as the calorimetric energy reconstruction because of the hadronic interactions. Above \SI{250}{MeV} of kinetic energy, there is a significant pion fraction that undergoes hadronic interactions, and for which the energy is underestimated. This can be clearly seen in the middle panel of Fig.~\ref{fig:resolution_all} as an increase in the relative energy bias. When comparing the calorimetric and \csda\ methods, the energy resolution is slightly better when using the range-based method, however the obtained resolution ranges from \SI{30}{\percent} to \SI{60}{\percent} below \SI{1}{GeV} of kinetic energy. New methods trying to fit the energy of tracks undergoing hadronic interactions in LArTPCs are currently being developed~\cite{dunecollaboration2024tracklengthextensionfittingalgorithm}, and it is foreseen to implement them in the future. For this paper, when needed for the neutrino direction reconstruction, the \csda\ method is used for the energy reconstruction of protons, and the calorimetric one for that of pions (as the method less-sensitive to mis-PID).


The situation is very similar for protons, as it can be seen in the right panel of Fig.~\ref{fig:resolution_all}. The two major differences from pion are: i) the energy regime in which hadronic interactions impair the energy resolution is higher ($>$\SI{500}{MeV}); ii) the gain in resolution obtained by using the \csda\ method is more significant. At low \Ekin, where most of the protons lie, the energy resolution is around \SI{15}{\percent} for the \csda\ method and \SI{20}{\percent} for the calorimetric method.




%% file: nu_eReco.tex
\subsection{Neutrino energy reconstruction}
\label{subsec:nuenergy}
In DUNE FD, both the energy of the lepton \Eereco\ or \Emureco{} (already presented in Secs.~\ref{subsubsec:erec_others} and ~\ref{subsubsec:erec-muons}, respectively) and of the hadronic system \Ehadreco{} can be reconstructed from the detected final state of a neutrino interaction. For the hadronic system, the reconstructed energy, \Ehadreco\, is calculated by adding all the registered hit energy depositions within the detector not associated with the tagged the lepton. This reconstruction assumes an average recombination factor for all the particles, and doesn't use any PID information except for muons, the only particle for which a different method is used for the energy reconstruction, as presented in Sec.~\ref{subsubsec:erec-muons}. In order to account for invisible energy within the detector that differs for the leptonic and hadronic components as well as for the neutrino nature, the reconstructed (anti-)neutrino energy is defined as: $E_{\nu}^{\rm reco} = \alpha E_{\rm lep}^{\rm reco} + \beta E_{\rm had}^{\rm reco}$, where $\alpha$ and $\beta$ are scaling coefficients obtained for each flavor sample by minimizing the average bias on the reconstructed energy and can be found in Table~\ref{tab:Escaling}.
While a more complex method applying different scalings based on the event topology or dependent on the energy would possibly yield slightly better apparent performance, it would ultimately be more subject to systematic uncertainties in a physics analysis. This more simple combination of \Ehadreco{} and \Elepreco{} was adopted here to get a quick estimation of the neutrino energy resolution. Note that however an actual physics analysis fit would probably prefer to rely on the individual \Ehadreco{} and \Elepreco{} observables without the need to combine them.
Neutrinos and anti-neutrinos of a given flavor are grouped together as they cannot be easily separated in the reconstruction.  In this section, the NC interactions are disregarded, as the neutrino only transfers a fraction of its energy, and focus on CC interactions only.

\begin{table} [!htb]
    \centering
    \begin{tabular}{|c|c|c|}
        \hline
         Sample & $\alpha$ (\Elepreco{} component) & $\beta$ (\Ehadreco{} component) \\
         \hline
         $\nu_e/\bar{\nu}_e$& 1.32  & 1.41 \\
         \hline
         $\nu_\mu/\bar{\nu}_\mu$& 1.08 & 1.47\\
         \hline
    \end{tabular}
    \caption{Fitted scaling coefficients applied to reconstruct the total neutrino energy with minimal bias}
    \label{tab:Escaling}
\end{table}

\subsubsection{Electron (anti-)neutrinos}
\begin{figure*}[tb]
    \centering
    \includegraphics[width=0.9\linewidth]{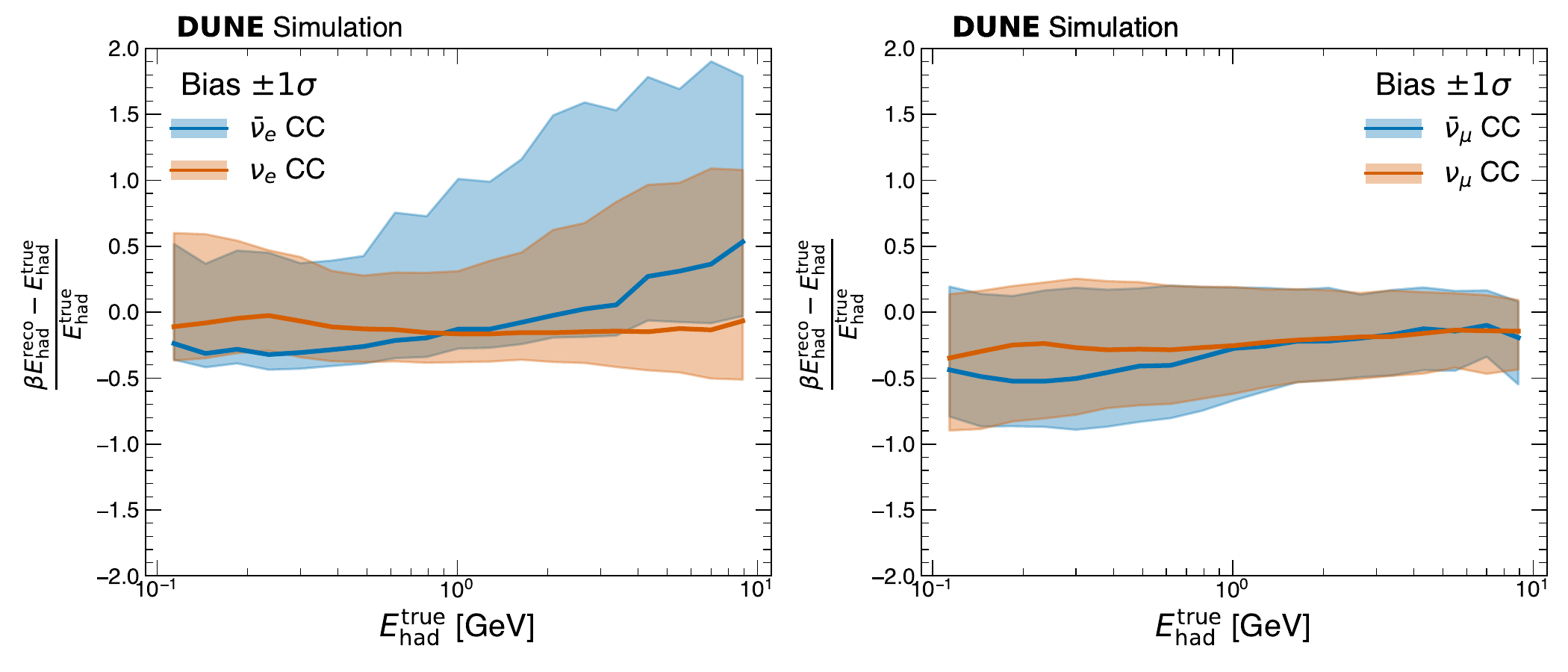}  
    \caption{Resolution and bias for the reconstructed hadronic energy \Ehadreco\, as a function of the true hadronic energy \Etruereco\ for: (Left) \nue\ and \nuebar\; (Right) \numubar\ and \numu{}.}
    \label{fig:nu-reco-had}
\end{figure*}

In the case of electron (anti-)neutrino interactions, the reconstruction of the energy is fully calorimetric. The scaled $\beta$\Ehadreco\, calculated as mentioned above,  is directly compared to the true energy of the hadronic system, $E_{\rm had}^{\rm true} = E_\nu^{\rm true} - E_e^{\rm true}$, in the left panel of Fig.~\ref{fig:nu-reco-had}. Two regimes can be noticed: one at relatively small hadronic energies, a region dominated by quasi-elastic interactions, where the resolution is the best ($\sim\SI{40}{\percent}$ below \SI{1}{GeV}); and a second one, at higher hadronic energies, where resonant and inelastic neutrino interactions, as well as inelastic hadronic interactions of the final state particles happen, leading to a loss in energy resolution ($>\SI{70}{\percent}$ above \SI{3}{GeV}). The resolution is also better for \nue\ than for \nuebar\ because the latter leads to the production of a primary neutron (which cannot be detected) instead of a proton.

\begin{figure*}[tb]
    \centering
    \includegraphics[width=\linewidth]{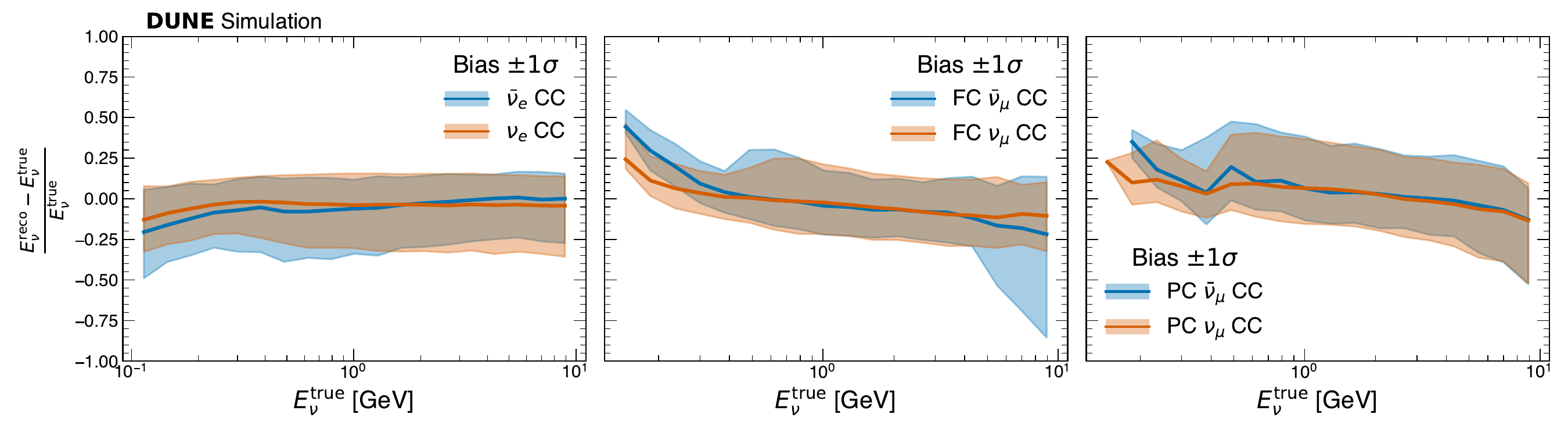}    
    \caption{Resolution and bias for the reconstructed neutrino energy \Enureco, as a function of the true neutrino energy \Enutrue\ for: (Left) \nue\ and \nuebar\; FC (Middle) and PC (Right), \numubar\ and \numu{}.}
    \label{fig:nu-ereco}
\end{figure*}

The left panel of Fig.~\ref{fig:nu-ereco} shows the same performance metrics but for the reconstruction of the neutrino energy, \Enureco{} obtained by linearly combining \Ehadreco\ and \Elepreco{}. The obtained resolution on \Enureco\ is below \SI{15}{\percent} at low energy, and around \SI{20}{\percent} at higher energies. Furthermore, the resolutions are very comparable for $\nu_e$ and $\bar{\nu}_e$, despite the worse performance obtained on the measurement of \Ehadreco\ for $\bar{\nu}_e$. This is directly due to the smaller transfer of energy to the hadronic system in the case of $\bar{\nu}$ interactions compared to $\nu$ interactions.

\begin{figure*}[tb]
    \centering
    \includegraphics[width=0.9\linewidth]{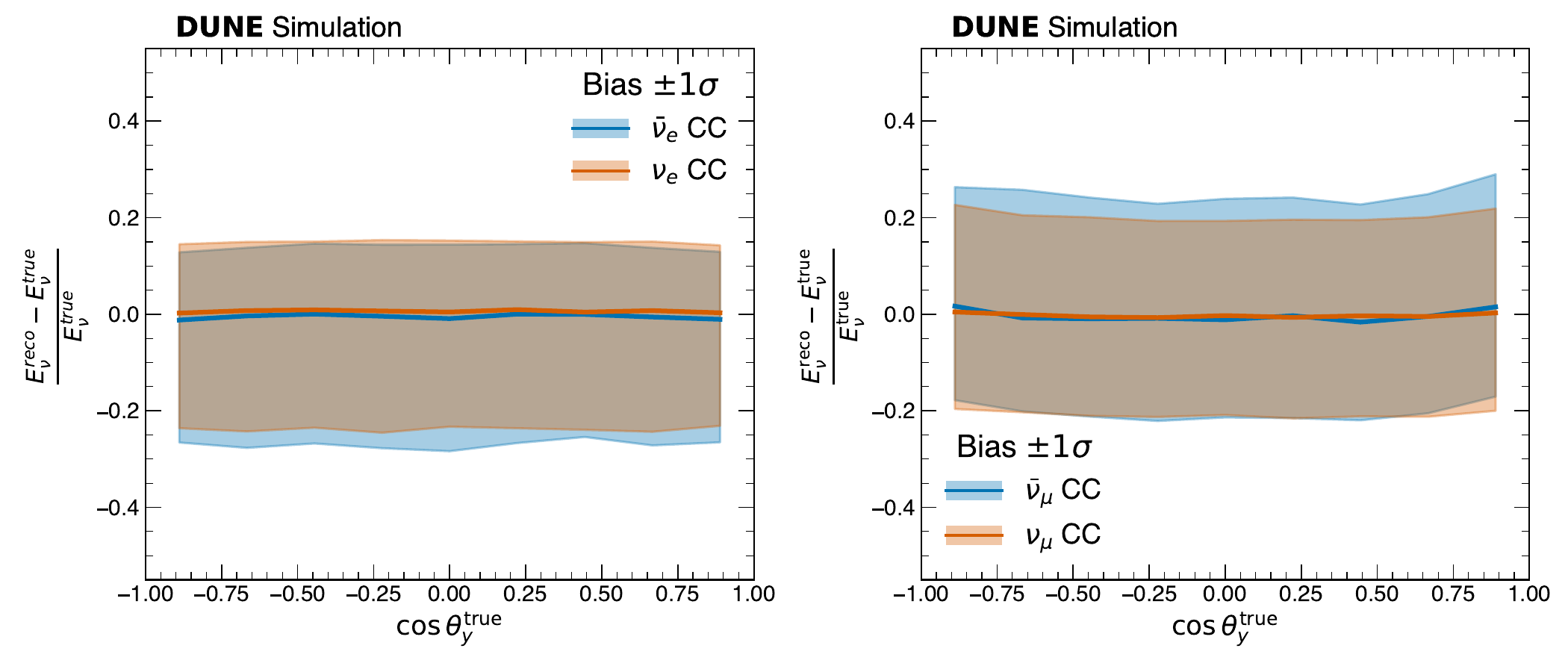}  
    \caption{Resolution and bias for the reconstructed neutrino energy \Enureco\ as a function of the neutrino direction along $y$, $\cos(\theta_y^{\rm true})$ for: (Left) \nue\ and \nuebar{}; (Right) \numu\ and \numubar{}.}
    \label{fig:nu-ereco-direc}
\end{figure*}

The energy reconstruction exhibits some dependency on the neutrino direction. The left panel of Fig.~\ref{fig:nu-ereco-direc} shows that the main effect is a smaller reconstructed energy for neutrinos traveling in a direction orthogonal to the $z$ direction, due to the arrangement of the plane wires and to the drift direction. This effect is however relatively small with a maximum absolute difference in the average energy bias between the orientations below \SI{3}{\percent}.

\subsubsection{Muon (anti-)neutrinos}

In the case of muon (anti-)neutrinos interactions, the reconstruction of the energy is relying on the \csda\ and \mcs\ methods for the muon part, and on the calorimetric measurement for the rest of the system. The results obtained when directly comparing $\beta$\Ehadreco\ to $E_{\rm had}^{\rm true} = E_\nu^{\rm true} - E_\mu^{\rm true}$ are shown in the right panel of Fig.~\ref{fig:nu-reco-had}: they exhibit similar behaviors as in the electron (anti-) neutrino case at low-energy (\SIrange{35}{40}{\percent} resolution below \SI{1}{GeV}). However, the obtained energy resolution on the hadronic system is significantly lower than in the electron neutrino case for multi-GeV neutrinos where it reaches $\sim\SI{30}{\percent}$. This is expected by the fact that the discrimination between the lepton and hadronic system is easier if the lepton is a muon, which leaves a long track in the detector, than when it is a electron, which can produce wide showers overlapping with the hadronic system.

\Enureco{} is again obtained by linearly combining \Ehadreco\ and \Elepreco\ and the obtained results are shown in the two right-most panels of Fig.~\ref{fig:nu-ereco}. The resolution on \Enureco\ is below \SI{10}{\percent} at low energy, and around \SI{15}{\percent} at higher energies for FC events. Furthermore, the resolutions are very comparable for \numu\ and \numubar, despite the worse performance obtained on the measurement of \Ehadreco\ for \numubar\ for the same reasons as above. Some difference can however be noticed on the bias of these two distributions at low and high energies due to an interplay between the differences in hadronic system and inelasticity distributions. As expected, events only partially contained don't allow to measure the energy in the most accurate way, especially for muons, hence leading to degraded neutrino energy resolutions for PC events. However the neutrino energy can still be measured with a resolution around \SI{20}{\percent} for multi-GeV PC \numu\ events.

The right panel of Fig.~\ref{fig:nu-ereco-direc} shows that the muon neutrino energy reconstruction is only moderately impacted by the incoming neutrino direction, with a slight loss of performance only noticed for neutrinos aligned with the drift direction $x$.


%% file: nu_dirReco.tex
\subsection{Neutrino direction reconstruction}
\label{subsec:nudirection}

\subsubsection{Method}
Three main algorithms have been developed to reconstruct the direction of the interacting neutrinos in the DUNE FD, which rely on different levels of information retrieved from the event reconstruction algorithms.

\begin{figure*}[!htb]
    \centering
    \includegraphics[width=0.45\linewidth]{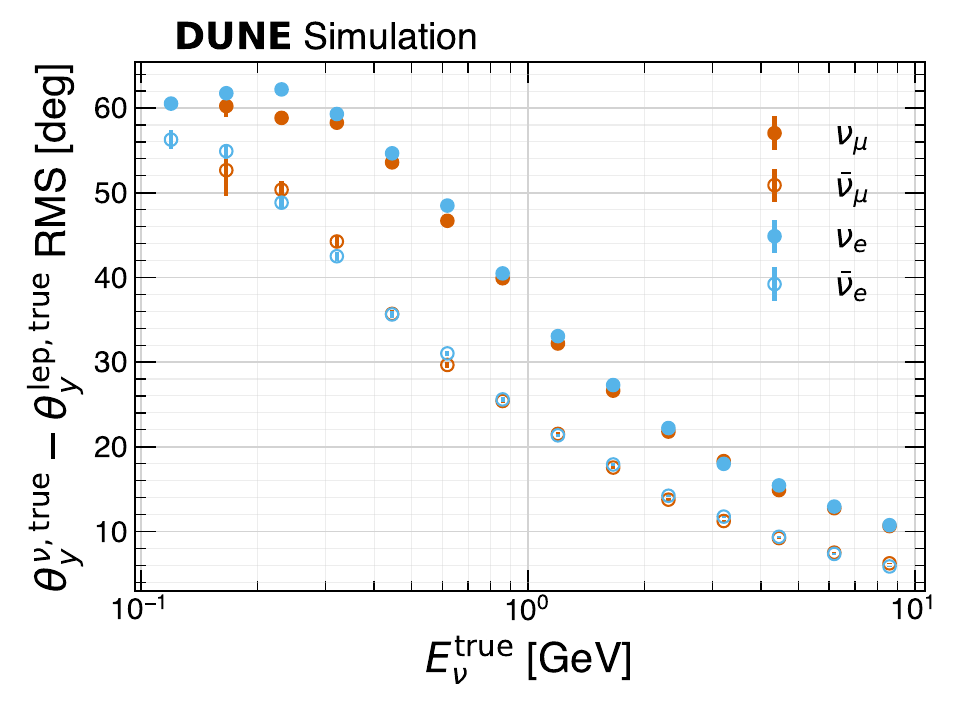}
    \caption{The RMS of the difference between the true neutrino and the true primary lepton directions along $y$, as a function of the true neutrino energy, for CC interactions.}
    \label{fig:nu-lep-angle}
\end{figure*}

\paragraph{Using just the reconstructed primary lepton.}
The simplest and most straightforward method for estimating the direction of the interacting neutrino uses only the information of the reconstructed primary lepton direction. The primary lepton particle is easy to reconstruct in DUNE FD, and as its direction tends to align with the neutrino's direction, as shown in Fig.~\ref{fig:nu-lep-angle}, one can directly use the reconstructed lepton direction as reconstructed neutrino direction. This is especially true for antineutrino interactions, which have a lower inelasticity (\ie, hadronic energy fraction) distribution than neutrino interactions. The obtained performance highly depends on the neutrino energy mostly because of Fermi motion: at lower energies the neutrino and target nucleon have similar momenta and the information on the neutrino direction gets heavily smeared out in the final state. There are intrinsic limitations of this method: i) it performs poorly for NC interactions; ii) it does not take into account the other reconstructed particles of the event for a more complete calculation. This method is typically the one applied by Water Cherenkov detectors for atmospheric neutrinos with an energy below a few \si{GeV} because of the Cherenkov threshold, which often suppresses the visibility of the heavier hadronic system compared to the lepton. In practice in DUNE, it just means picking the most energetic reconstructed particle to be the lepton in the event.

\paragraph{Using all the reconstructed particles.}
A second method to estimate the direction of the interacting neutrino is possible if the hadronic system can be reconstructed: it relies on the use of all reconstructed final state particles and summing their individual momenta, which implies the determination of their type/ID, energy, and direction. Reaching enough accuracy on these three metrics allows to improve the estimation of the neutrino direction with respect to the lepton-only method. The caveat of this method is that it can only be applied to events for which particles have been reconstructed.

\begin{figure*}[!htb]
    \centering
    \includegraphics[width=0.45\linewidth]{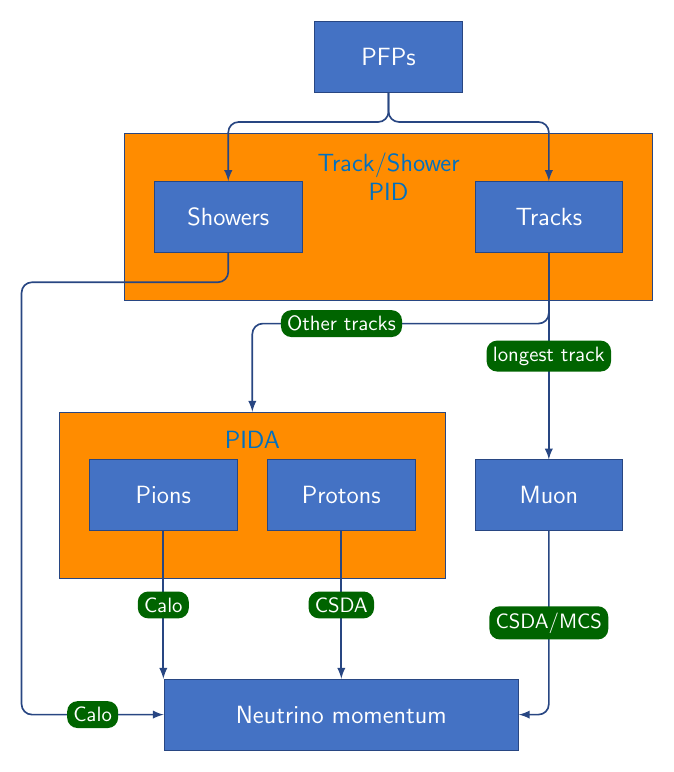}
    \caption{Flowchart of the PID selection performed, and the methods used, to estimate the momentum of every of the final state particles when calculating the neutrino direction using all particles reconstructed in the event. The \emph{Calo} refers to the method (calorimetric) deployed to estimate the kinetic energy of individual particles.}
    \label{fig:angle-pid}
\end{figure*}

A flowchart of this method is shown in Fig.~\ref{fig:angle-pid}. It starts with all PFP candidates, identifies their type, estimates their total momentum based on either the \csda, \mcs, or $\rm{Calo}$ (calorimetric estimation of the kinetic energy) method, which in the end is combined with their reconstructed direction to obtain the particles' 3-momentum. Once the 3-momentum of all the final state particles $\vec{p}_i$ is obtained, the reconstructed neutrino direction $\hat{p}$ is obtained as:

\begin{equation}
\hat{p} = \frac{\sum_i{\vec{p}_i}}{\left\lVert\sum_i{\vec{p}_i}\right\rVert}
\end{equation}
As shown in the flowchart, the momentum of muons and protons is estimated by using the CSDA method as it allows to reach the best performance. However, the charged pions' momenta are estimated using the calorimetric measurement of the energy depositions. This choice was made as this reconstruction method is less biased by pion inelastic scattering interactions.

\paragraph{Using only calorimetric hits.}
A third method focuses on the determination of the neutrino direction using only the low-level reconstructed hit information in the 3 detector planes ($u,v,w$). The major advantage of this method is that it does not rely on any high-level particle-reconstruction information, and it can thus serve as a fallback method for any scenario in which the direction of an interacting neutrino is needed. It relies solely on the deposited energy, identifying the direction of the kinetic energy deposition to the momentum direction, without taking into account the various particle masses. The flow of this reconstruction method is the following:

\begin{enumerate}
    \item For each of the 3 2D views (U, V and W), the reconstructed 2D hits are projected in the view coordinate system built from a normal vector to the wire direction and the time axis (associated to the drift direction). The hit coordinates in each view are shifted so that the reconstructed vertex lies at the origin.
    \item The average 2D direction for each view ($\alpha^\text{view}$) is calculated in the complex plane by averaging the direction to each hit from the origin (angle $\alpha_i$ with the time axis), weighted by the hit reconstructed charge ($q_i$): $\alpha^\text{view} = \arg{\left(\sum_i{q_i e^{j\alpha_i}}\right)}$ where $j$ is the imaginary unit.
    This hit reconstructed charge is corrected by the drift attenuation as the drift time is known for each hit. However no recombination correction is applied as the \dedx information is inaccessible at the hit level.
    \item The 3D average energy direction is fitted from the 2D measurements by building a $\chi^2$ from the differences between the observed 2D angle and projected 3D direction in each view. Indeed, only 2 independent measurements would be necessary to reconstruct a 3D direction, and the fitting procedure allows to find the best compromise between the 3 existent plane measurements.
\end{enumerate}

\begin{figure*}[tb]
    \centering
    \includegraphics[width=0.9\linewidth]{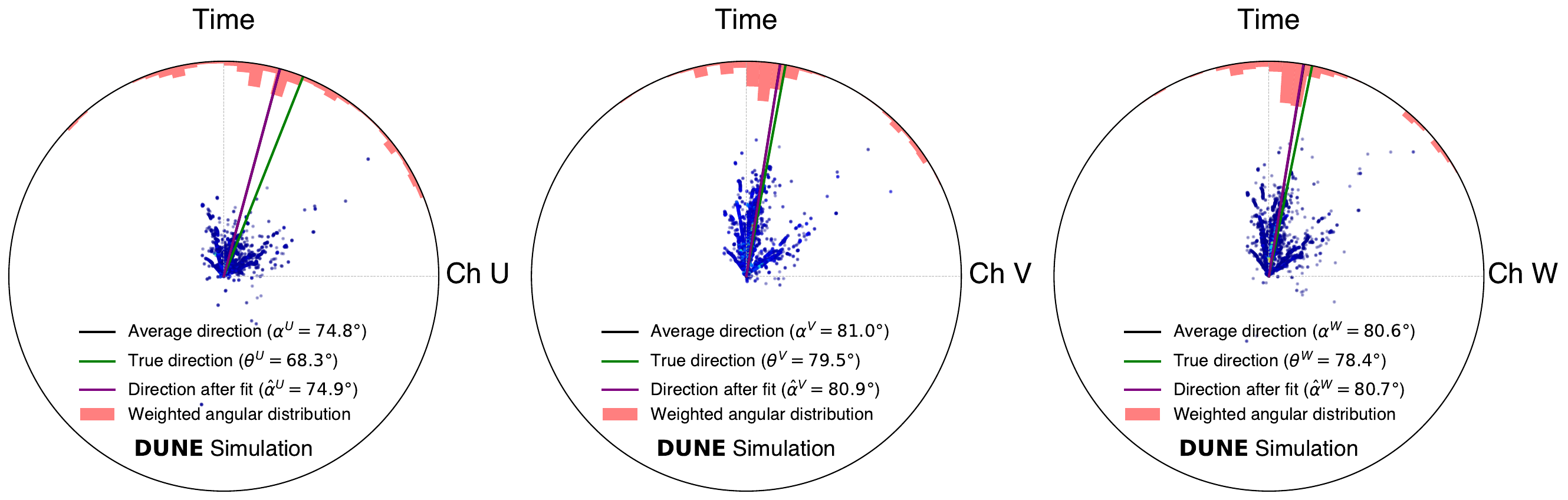}
    \caption{Example of the calorimetric based direction reconstruction showing the direction estimation for a simulated NC interaction with \SI{12}{GeV} of visible energy. This higher energy interaction was selected for illustration purpose even though it is not used in the analysis.}
    \label{fig:calo-angle-example}
\end{figure*}

This whole process is illustrated in Fig.~\ref{fig:calo-angle-example} where the true neutrino direction projected on the 2D planes ($\theta^\text{view}$) as well as the reconstructed neutrino direction after the $\chi^2$ fit projected on the 2D planes ($\hat\alpha^\text{view}$) are shown for comparison.

\subsubsection{Results}

\begin{figure*}[tb]
    \centering
    \includegraphics[width=\linewidth]{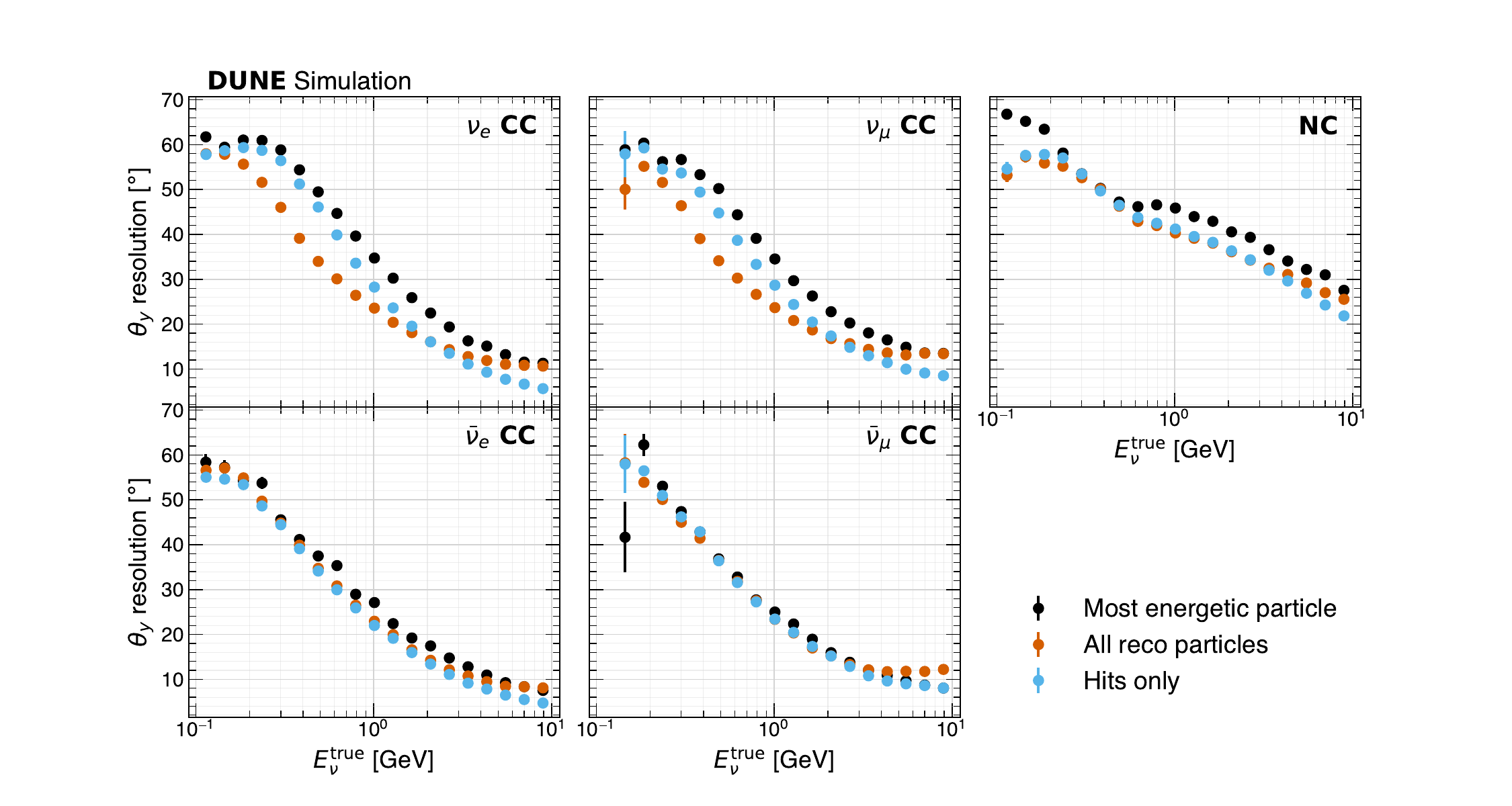}
    \caption{Resolution (in degrees) of the reconstructed particle (Top) and antiparticle (Bottom) direction, as a function of their true energy for \nue\ (Left) and \numu\ (Middle) CC interactions, and NC interactions (Right).}
    \label{fig:angle-reco-with-e}
\end{figure*}

All the resolutions hereafter are shown for \tetay{}, the zenith angle (angle of the neutrino with the upward $y$ direction) and are calculated with the quartiles method applied on the absolute difference in the reconstructed \tetay{} angles. All the results discussed (and figures shown), referred for convenience as for neutrino interactions, are for particles and antiparticles of the same flavor combined, as it was checked that the performance in the two cases is comparable.

The obtained angular resolutions with the three methods for all the $\nu_e$ CC, $\nu_\mu$ CC, $\bar{\nu}_e$ CC, $\bar{\nu}_\mu$ CC and NC interactions are presented in Fig.~\ref{fig:angle-reco-with-e}, which shows that the angular resolution improves with energy for all the methods, as expected by the relative decrease of the nucleon Fermi momentum with respect to the neutrino energy. At low energy, the momentum of the nucleon inside the argon nucleus, on which the neutrino interacts, is relatively high  with respect to the momentum of the neutrino itself. As a consequence, the momentum of the final state particles is equal to the sum of the momentum of the incoming neutrino and of the target nucleon which leads to a smearing of the initial neutrino direction. It thus becomes impossible to accurately reconstruct the neutrino direction at low enough energies. Figure~\ref{fig:angle-reco-with-e} also shows that the method using all the reconstructed particles systematically outperforms the method using only the most energetic particle, demonstrating the advantage of reconstructing the hadronic system that is possible in DUNE. However, these two methods converge at higher energies as the reconstruction deteriorates and tends to cluster together all the hits into a single large shower. This improvement is much stronger for neutrinos compared to anti-neutrinos given that the visible energy in the hadronic system is much lower for anti-neutrinos due to the presence of a primary neutron in the final state (\vs\ a primary proton for neutrinos) and the lower interaction inelasticity. This lower inelasticity also has the effect of improving the correlation between the lepton and neutrino direction, hence leading to very similar performance overall for the best reconstruction method for neutrinos and anti-neutrinos at a given energy.
As expected, the hits-only method underperforms the all-reconstructed particles at lower $<\SI{1}{GeV}$ energies, and outperforms for higher $>\SI{2}{GeV}$ energies for neutrinos. Indeed, at lower energies the missing information about the relative masses from the particles impairs the performance of the hits-only method, while this effect is reduced at higher energies where it is compensated by the easier handling of complex topologies by removing the need for a performant high-level reconstruction of tracks and showers. The performance for NC interactions are worse than for CC interactions, as expected given that the neutrino only transfers part of its momentum in the interaction. For these interactions, the hits-only methods is the one that performs in average the best as expected by the fact that there is no primary lepton to reconstruct and several small energy deposits not forming tracks or showers.

\begin{figure*}[tb]
    \centering
    \includegraphics[width=\linewidth]{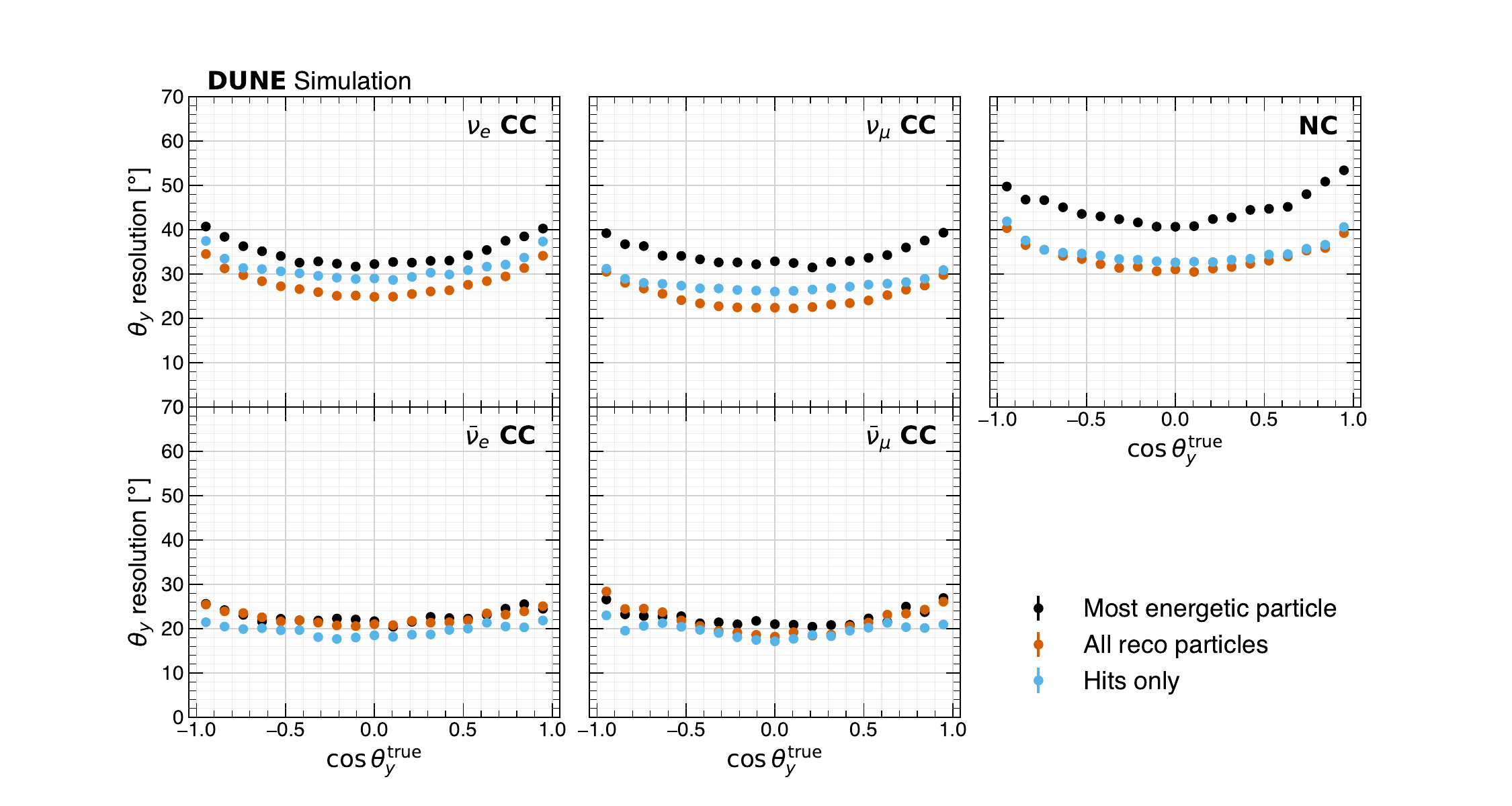}
    \caption{Resolution of the reconstructed particle (Top) and antiparticle (Bottom) direction, as a function of the true neutrino direction with respect to the zenith for \nue\ (Left) and \numu\ (Middle) CC interactions, and NC interactions (Right).}
    \label{fig:angle-reco-with-dir}
\end{figure*}

The same resolutions are shown as a function of the (anti-)neutrino direction with respect to the zenith direction ($\cos\theta_y$) in Fig.~\ref{fig:angle-reco-with-dir}. All the methods display the same behavior as a function of the neutrino direction: the best resolutions are obtained for the neutrinos orthogonal to the zenith direction while the most degraded resolutions are obtained for up-going and down-going neutrinos. This can be once again explained by the alignment of the vertical direction with the collection plane wires, inducing a loss of resolution for particles traveling in that direction.

\begin{figure*}[tb]
    \centering
    \includegraphics[width=\linewidth]{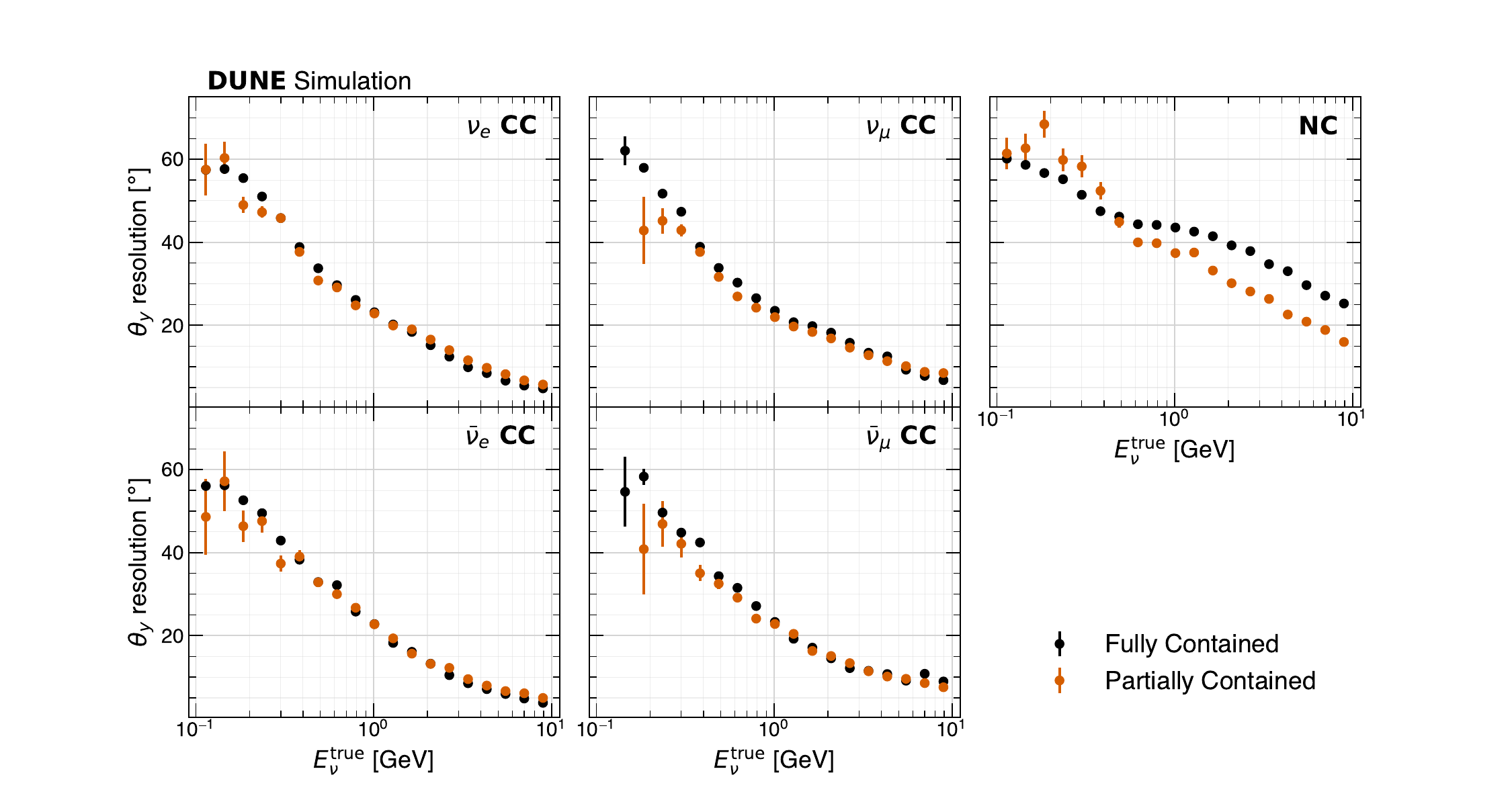}  
    \caption{Resolution (in degrees) of the reconstructed particle (Top) and antiparticle (Bottom) direction as a function of the true neutrino energy for \nue\ (Left) and \numu\ (Middle) CC interactions, and NC interactions (Right), using the hybrid reconstruction method.}
    \label{fig:direc-res-etrue-combined}
\end{figure*}

A study was carried out on the direction resolution as a function of the calorimetric reconstructed neutrino energy for events with a single and more than one reconstructed particle. The results were used to build a hybrid direction reconstruction procedure, that applies one of the three mentioned methods depending on the number of reconstructed particles and the reconstructed neutrino energy. The hits-only method was found to be the best in almost all cases where a single particle is reconstructed, therefore chosen to be used in the hybrid model for these cases. For CC events in which multiple particles are reconstructed, an applicability threshold of \SI{1.3}{GeV} in reconstructed calorimetric energy is decided: below it, the method based on reconstructed particles is applied, and the hit-based method above it. Similar events tagged as NC will always use the hit-based method. The results obtained with this hybrid method are shown in Fig.~\ref{fig:direc-res-etrue-combined}. It can be seen that the performance is rather independent of the full containment of the interaction, except for NC interactions for which the containment and transferred energy are very intertwined. Additionally the obtained performance is very similar between neutrinos and anti-neutrinos. For CC interactions, the obtained performance is a resolution around \SI{25}{\degree} at \SI{1}{GeV} that goes down to \SI{7}{\degree} for \nue\ and \SI{9}{\degree} for \numu\ at \SI{10}{GeV}. The counterintuitive slightly better performance obtained for PC events across part of the energy range can be understood by the interplay between the lepton containment and the inelasticity of the interaction which impacts the direction reconstruction. The NC interactions feature a degraded resolution as expected, but this should still be sufficient to help in an oscillation analysis, by setting total flux constraints.

%% file: nu_flavReco.tex
\subsection{Neutrino flavor determination}
\label{subsec:nunubar}
A convolutional visual network (CVN) has been deployed to perform image recognition with the goal of providing neutrino interaction classification in DUNE's FD modules. This kind of network is particularly suited for DUNE given the nature of the detector output that can be represented as a set of three 2D images of the neutrino interaction (one for each of the $u, v, w$ planes). This CVN proved~\cite{cnn-beam} to be highly efficient and shows high purity in the selections of electron and muon neutrino interactions on samples of beam neutrino events. After it was retrained on a sample of atmospheric neutrinos, the same CVN was deployed to assess its performance on atmospheric neutrino events.

The CVN model uses as input three 300x300 pixel images/maps (one for each plane of the LArTPC), which are generated from the reconstructed 2D hits on the readout wires. A pre-selection criterion is applied to each neutrino event used for training: it must contain at least 100 hits, the neutrino energy must be less than \SI{20}{GeV}, and the true neutrino interaction vertex has to be within the detector's FV (as defined in Sec.~\ref{sec:anaSetup}). This pre-selection focuses the analysis on \nue\ CC, \numu\ CC, and NCs interactions, while excluding \nutau\ for which a few interactions only are expected per module every year. After pre-selection, a total of \SI{1.84}{M} pixel maps were used for training, \SI{80}{k} for validation and \SI{80}{k} for testing. 

\begin{figure*}[tb]
    \centering
    \includegraphics[width=0.9\linewidth]{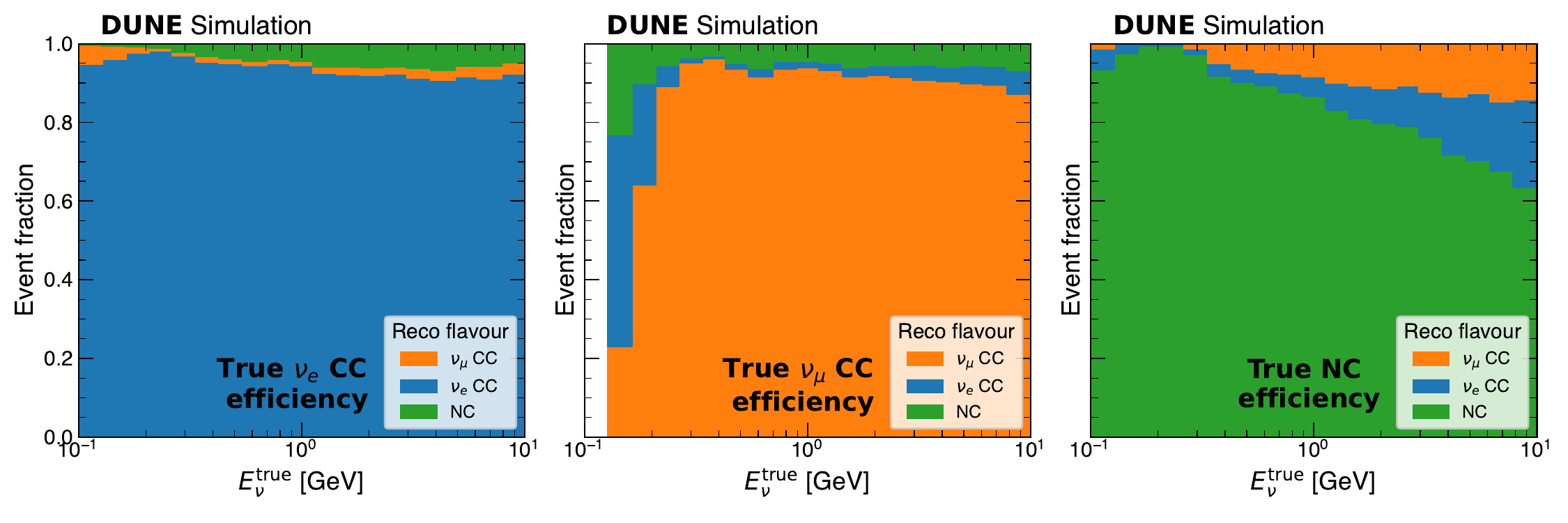}
    \caption{Decomposition of true neutrino interaction types as reconstructed CVN types, for \nue\ CC (Left), \numu CC (Middle), and  NC (Right), as a function of the true neutrino energy.}
    \label{fig:cvn-true}
\end{figure*}
\begin{figure*}[tb]
    \centering
    \includegraphics[width=0.9\linewidth]{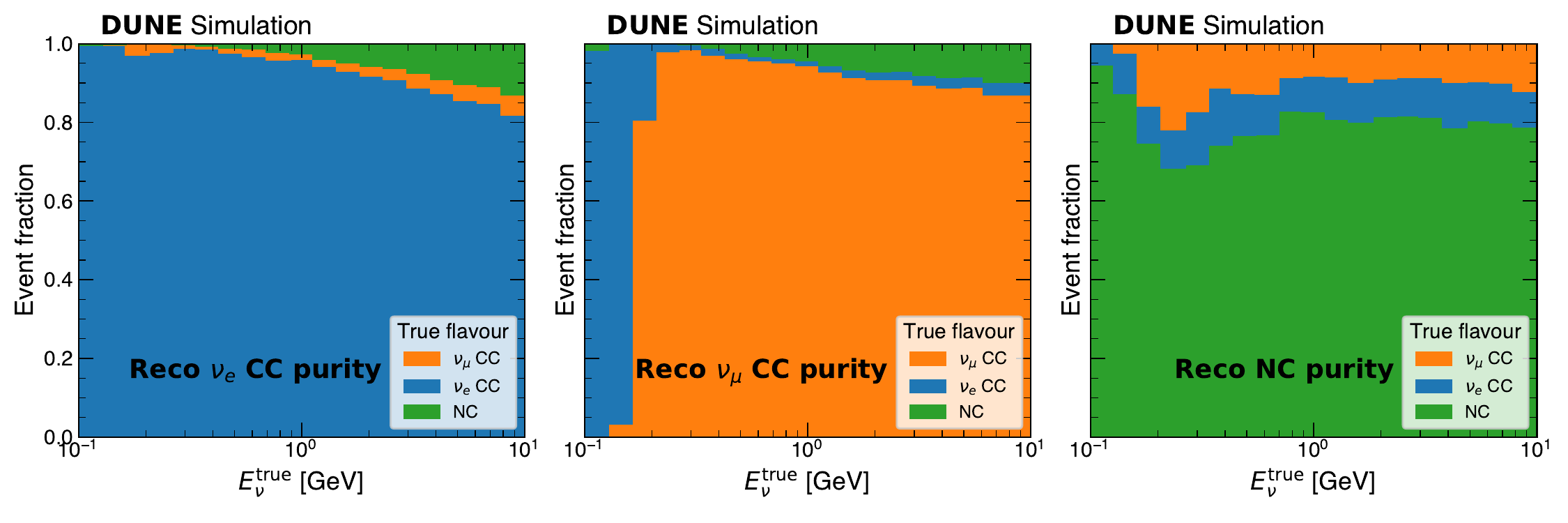}
    \caption{Decomposition of reconstructed CVN types as true neutrino interaction types, for \nue\ CC (Left), \numu CC (Middle), and  NC (Right), as a function of the true neutrino energy.}
    \label{fig:cvn-reco}
\end{figure*}

The performance is assessed on the full selected dataset according to the selections described in Sec.~\ref{subsec:fiducialization} For every event, the reconstructed flavor is selected as the flavor with the highest output score from the CVN.
The obtained results are shown as stacked histograms, split by true labels in Fig.~\ref{fig:cvn-true}, and split by reconstructed labels in Fig.~\ref{fig:cvn-reco}.
The selection efficiency for \numu\ and \nue\ CC (NC) interactions is above \SI{90}{\percent} (\SI{60}{\percent}) across the energy range, as can be seen in Fig.~\ref{fig:cvn-true}. The efficiency in selecting the CC interactions is relatively flat with the energy while there is a continuous decrease of the selection efficiency as a function of the energy for NC interactions. This can be understood by the fact that high-energy NC interactions might have similar features as CC interactions with many particles in the final state.
This very same effect leads to a drop of the selection purity with the energy, as shown in Fig.~\ref{fig:cvn-reco}. 
The lower witnessed purities and efficiencies for $\nu_\mu$ CC interactions below \SI{200}{MeV} can be explained by the low number of $\nu_\mu$ able to undergo CC interactions due to phase space suppression, as well as by the low energy of the produced muon. The interaction containment was tested to have a minimal impact on the performance, once disentangled from the amount of transferred momentum.

\begin{figure*}[tb]
    \centering
    \includegraphics[width=0.9\linewidth]{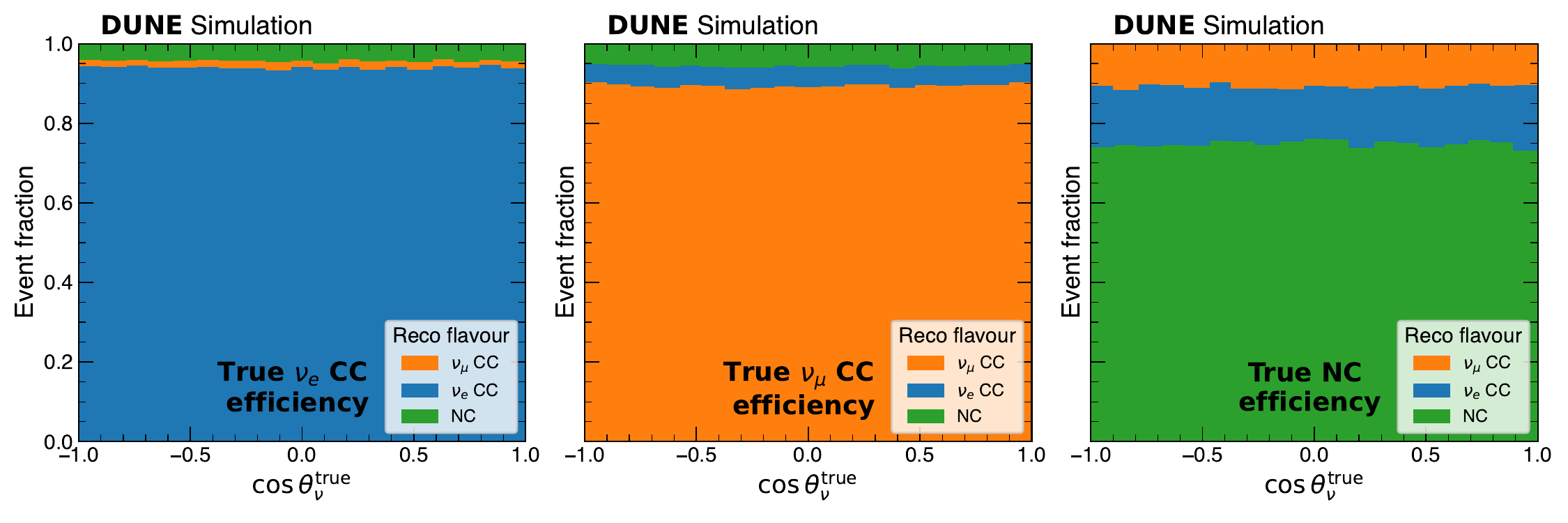}
    \caption{Decomposition of reconstructed CVN types as true neutrino interaction types, for \nue\ CC (Left), \numu CC (Middle), and  NC (Right), as a function of the true neutrino angle with the zenith direction ($\cos(\theta_y^{\rm true})$).}
    \label{fig:cvn-with-cz}
\end{figure*}

Figure~\ref{fig:cvn-with-cz} demonstrates that the performance from the CVN is rather independent from the direction of the neutrino. The overall obtained performance is comparable to what was previously described for beam events~\cite{cnn-beam}, with some slight loss in performance as the pattern recognition becomes intrinsically more difficult when neutrinos can arrive from any direction, with a difficult generalization to the multiple incoming directions, and with a wider energy flux.


%% file: Summary.tex
\section{Summary and discussions}

\subsection{Vertex}
The current DUNE vertex reconstruction algorithm for atmospheric neutrino events: 
\begin{itemize}
    \item provides a sub-cm level position resolution over a wide energy range; 
    \item has an efficiency (\ie, the fraction of events with sub-cm resolution) depending on the neutrino energy, between \SIrange{75}{90}{\percent} for CC fully contained events;
    \item has a strong energy and neutrino direction dependence of the efficiency on the $y$ axis, understood to be caused by the FD design (\ie, wires pitch and orientation in the planes of the APAs).
\end{itemize}

The performance (resolutions and efficiency of reconstructing a vertex  within \SI{1}{cm} of the true vertex) is similar to that obtained for the beam sample~\cite{DUNE:2022VTX}. The efficiencies are higher than 80\% across all energies and zenith angles in $x$ and $z$ directions. The lowest efficiency is found for the $y$ direction (the direction along which the collection plane's wires are oriented along, therefore providing less spatial resolution), where the efficiency drops to $\sim76\%$ at low-E ($<$\SI{1}{GeV}) and all zenith angles. To improve even further these efficiencies, a possible approach is to modify in some way the current network: either modify its architecture/settings or change the type of the network itself (\eg, use a graph neural network instead or in combination with the present U-ResNet).


\subsection{Particle reconstruction}
The \textit{particle reconstruction efficiency} depends on both the energy and the type of the particle:
\begin{itemize}
    \item for muons and electrons, the reconstruction efficiency is above 90\% and rather flat from $\sim$\SI{.6}{GeV} up to \SI{10}{GeV} momentum range;
    \item protons (pions) start with an efficiency around 90\% (80\%) between $\sim$\SI{.6}{GeV} and $\sim$\SI{1}{GeV}, which decreases steadily up to 60\% at $\sim$\SI{10}{GeV} after; 
    \item for photons, the efficiency as a function of momentum is between 65\% and 85\% from \SI{0.2}{GeV} to $\sim$\SI{1}{GeV}, increasing after to approach $\sim$100\% at $\sim$\SI{10}{GeV}. 
\end{itemize}

While the best resolution for the \textit{particle direction reconstruction} is obtained for muons, which have a resolution below \SI{5}{\degree} at a kinetic energy (\Ekintrue) of \SI{1}{GeV}, similar trends can be observed for all particle types:
\begin{itemize}
    \item the resolution is relatively poor at low-\Ekintrue{} (below \SI{100}{MeV}), in particular for proton tracks, which span $\mathcal{O}(cm)$ at these energies;
    \item the best resolution for shower-like particles is obtained for \Ekintrue{} between \SI{100}{MeV} and \SI{1}{GeV};
    \item at higher \Ekintrue{} the resolution degrades for shower-like particles, as events become more crowded, and the accuracy in the reconstruction of the individual particles decreases.
\end{itemize}


The \textit{particle energy resolution} depends on the particle-type:
\begin{itemize}
    \item Muons: DUNE can measure the contained-muons' energy with $\sim$2\% resolution with the \csda\ method. For uncontained muons, the energy can be estimated through a MCS measurement, with the obtained resolutions highly dependent on the contained track-length and slightly on the track direction;
    \item Charged pions: out of the two methods studied, the CSDA performs better (but can fail catastrophically in the case of bad PID). The obtained $>\sim30\%$ resolution is limited by inelastic hadronic interactions. 
    \item Protons, compared to pions, exhibit a few differences: the hadronic interactions manifest at higher \Ekintrue{} values ($>1$\,GeV); there are more significant gains with the CSDA method.
    There is good resolution at low-\Ekintrue: $<15\%$ below \SI{400}{MeV};
    \item Electrons: at low-\Ekintrue{} ($<400$\,MeV), the resolution is limited by shower particles detection thresholds (soft electrons); at mid-\Ekintrue{} ($0.4-1.2$\,GeV), the resolution is better than 10\%; at high-\Ekintrue{} ($> 2$\,GeV), the resolution degrades because of over-clustering of hits. 
\end{itemize}

\textit{Particle identification} in DUNE reaches a good separation between track-like and shower-like particles by selecting track-like particles with an efficiency of \SI{89}{\percent} while rejecting \SI{91}{\percent} of the showers. A particle identification is also performed over the tracks, allowing to tag protons by using a simple PIDA technique (\dedx\ profile): they can be selected with an efficiency (purity) of $\sim$87\% (91\%). An accurate discrimination between charged pions and leptons can be achieved only when/if additional criteria (\eg, the track scattering to the track length) are considered.

\subsection{Neutrino energy reconstruction}

The reconstructed neutrino energy
\begin{itemize}
    \item has a resolution that varies between \SI{10}{\percent} and \SI{20}{\percent} depending on the event flavor, energy, and containment; 
    \item depends on the true energy due to containment and interaction topologies; 
    \item depends loosely (sub-leading order) on the neutrino direction. 
\end{itemize}

\begin{figure*}[tb]
    \centering
    \includegraphics[width=0.45\linewidth]{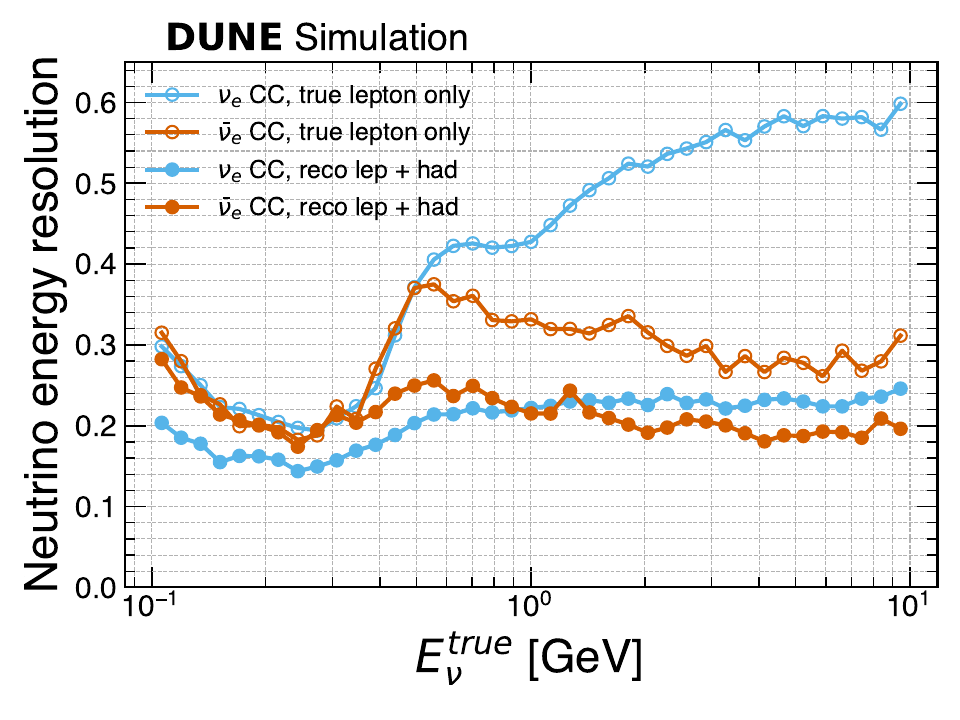}  
    \includegraphics[width=0.45\linewidth]{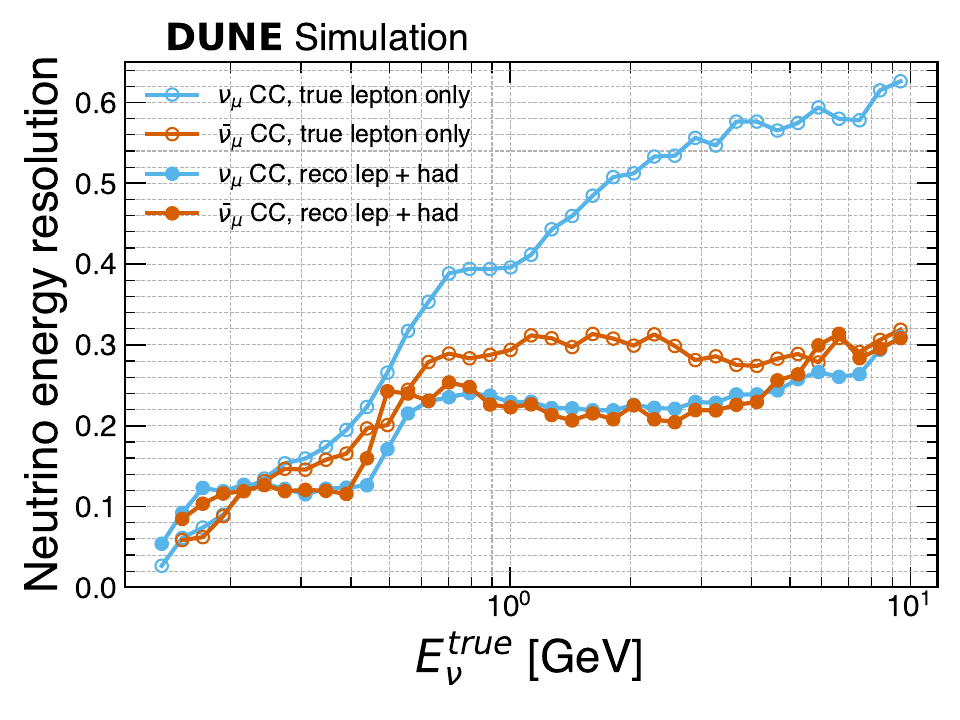}  
    \caption{Neutrino energy resolution as a function of the true neutrino energy for \nue/\nuebar\ (Left) and \numu/\numubar\ (Right), using the true lepton energy on the one hand, and the reconstructed leptonic/hadronic energies on the other hand.}
    \label{fig:energy-had-comp}
\end{figure*}

The current energy reconstruction is relatively simple and relies heavily on the estimation of the hadronic energy through the measurement of the total deposited charge. Using PID information for the energy reconstruction should allow to improve these resolutions as demonstrated in~\cite{lartpc-perfs} by using more precise energy reconstruction methods such as CSDA. Additionally, using the 3D reconstruction of the track-like particles, instead of the 2D hits calorimetry, would allow to apply recombination corrections matching the measured \dedx\ profile of each particle instead of an average recombination factor. However, despite the imperfections of the determination of the hadronic energy in the current reconstruction, the information it brings is already enough to considerably improve the energy resolution with respect to only using the primary lepton information. This is clearly demonstrated in Fig.~\ref{fig:energy-had-comp}, that shows a direct comparison of the energy resolution performance in two cases:
\begin{itemize}
    \item The true lepton energy is used and rescaled at best to be unfolded into a neutrino energy by minimizing the bias, thus representing the resolution obtained with an idealized detector only accessing the lepton information;
    \item The reconstructed neutrino energy from the reconstructed hadronic and leptonic quantities as presented in Sec.~\ref{subsec:nuenergy}.
\end{itemize}

At lower energies (\SIrange{100}{400}{MeV}), there is no improvement on the energy resolution brought by the detection of the hadronic system for anti-neutrinos as expected by the fact that it mainly consists of a primary neutron which is undetected. For \nue\ there is however some gain from \SIrange{5}{10}{\percent} on the absolute resolution with the detection of the hadronic system, mainly through the detection of the primary proton. A similar effect can be witnessed for \numu\, with an additional small decrease in performance at very low energies ($<$\SI{200}{MeV}), possibly due to the complexities introduced by very low-energy muons. At higher energies, the gains in resolution become more obvious. The absolute neutrino energy resolution is for example halved going down from $\sim\SI{40}{\percent}$ to $\sim\SI{20}{\percent}$ around \SI{1}{GeV} for \nue. A very important gain is also seen for \nuebar\ even though slighly less important as the share of the energy going into the hadronic system is smaller for anti-neutrinos. For \numu\ and \numubar\ similar conclusions can be drawn. The main difference being that at higher energies, around \SI{10}{GeV}, using only the true lepton information yields similar resolutions as using the full reconstructed system for \numubar. This is because using the true muon energy has the same effect as measuring the hadronic system. But at those energies, the muon is uncontained, and its energy is imperfectly estimated through the MCS method.

\subsection{Neutrino direction reconstruction}
Three direction reconstruction methods have been developed and studied for the atmospheric analyses: using lepton-only information, using all reconstructed particle, and using just reconstructed hits.
\begin{itemize}
    \item At low-E ($<$\SI{1}{GeV}) in CC interactions, using all reconstructing particles provides the best resolution;
    \item At high-E, in CC interactions, a pointing resolution $\sim5^{\circ}-10^{\circ}$ is obtained at \SI{10}{GeV} with the hits-only method;
    \item For NC interactions, some pointing capabilities exist, $\sim20^{\circ}$ at high-E and $\sim 40^{\circ}$ at \SI{1}{GeV};
    \item The direction resolution is dependent on the direction itself for all the methods.
\end{itemize}

The poor resolutions obtained at low energy can hardly be mitigated by any reconstruction means, because of the irreducible Fermi motion. However, the resolution limit has not been reached yet, and some newer methods relying on machine learning techniques (\eg, CNNs) might prove to be more efficient at reconstructing the neutrino direction.

\begin{figure*}[tb]
    \centering
    \includegraphics[width=.9\linewidth]{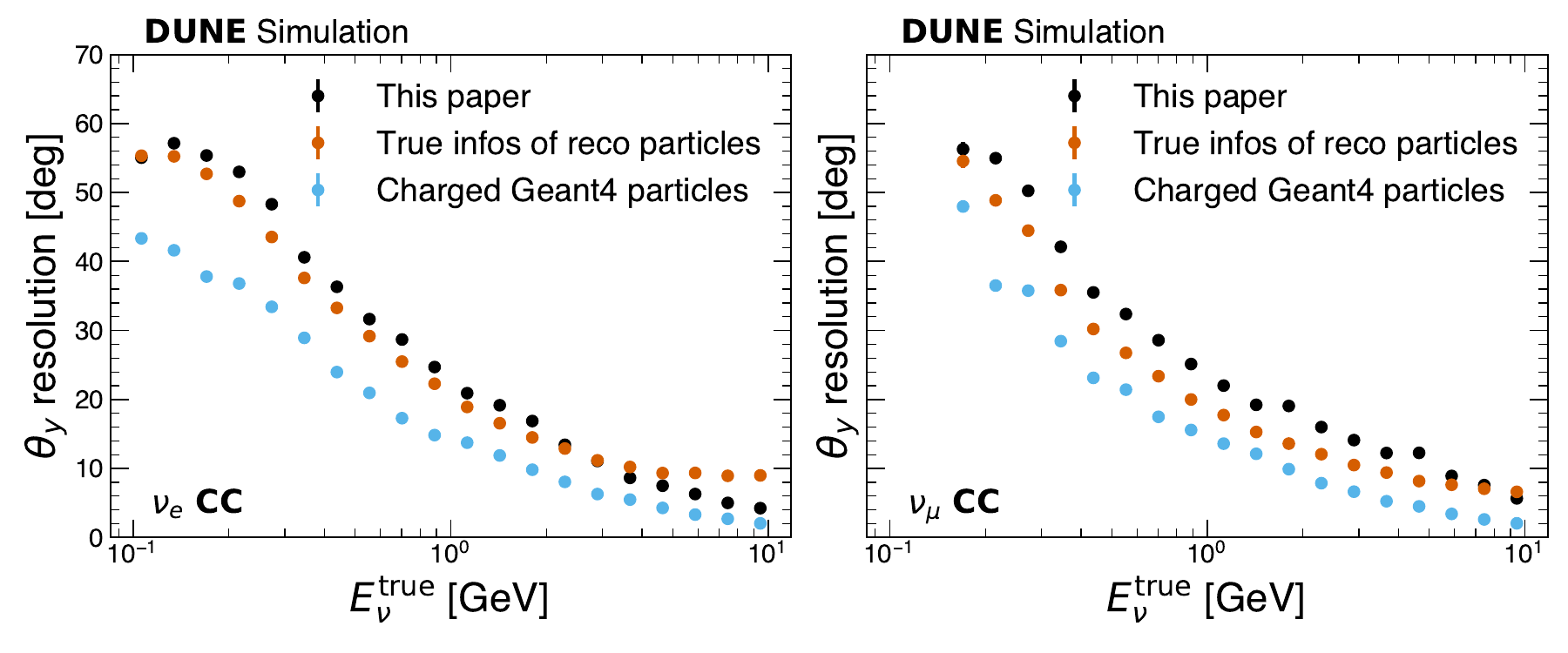} 
    \caption{Resolution in degrees on the reconstructed neutrino direction as a function of the true neutrino energy for \nue\ (Left) and \numu\ (Right) CC interactions using the hybrid reconstruction method, for three cases: current implementation, when the 3-momentum of every used Pandora reconstructed particle is perfectly determined, and for the case in which all true final state particles (except the neutrons) are perfectly reconstructed.}
    \label{fig:perfect-angle-reco}
\end{figure*}

To visualize what are the physical boundaries for any future improvements, Fig.~\ref{fig:perfect-angle-reco} compares the neutrino direction resolution obtained with the hybrid-method for three settings: current implementation, when the 3-momentum of every used Pandora reconstructed particle is perfectly determined, and for the case in which all true final state particles (except the neutrons) are perfectly reconstructed. First, it can be seen that improving the current PID and energy reconstruction alone, without improving the lower-level reconstruction of Pandora (\eg, hits, clusters, PFP, \etc), would only provide a relatively small improvement in neutrino direction resolution. Second, Fig.~\ref{fig:perfect-angle-reco} shows that an ideal particle reconstruction (estimated by using all primary Geant4 particles true momentum, except the neutrons) could only provide a sizable increase in performance below \SI{1}{GeV}, being already close to the current performance at higher energies. The limit is set by the inability to detect neutrons and the irremediable presence of nuclear effects. So at best, any sophisticated machine learning method could only try to approach this limit. However, ongoing development of Pandora is looking to deliver improvements in the lower-level reconstruction.

\subsection{Neutrino flavor determination}
An overall good performance of the CVN was obtained, with $\sim$\SI{90}{\percent} selection efficiency at tagging the neutrino flavour, over the whole 0.1--10\GeV\ energy range studied for CC interactions, with a purity above \SI{90}{\percent} below \SI{1}{GeV} and above \SI{80}{\percent} up to \SI{10}{GeV}. No significant bias of the CVN with respect to the zenith angle was observed: the selection purities are similar for all directions. Overall, the retrained-CVN performance is similar to that obtained in DUNE on the beam samples~\cite{cnn-beam}.

%% file: Outlook.tex
\section{Conclusions}


In this paper, the reconstruction performance of atmospheric neutrinos interacting in the LArTPC horizontal drift module of the DUNE Far Detector has been presented. For the first time in DUNE, a full simulation and reconstruction chain has been deployed to study atmospheric neutrinos with energies between \SI{0.1}{GeV} and \SI{10}{GeV}. The Pandora pattern-recognition software package was used for vertex and particle reconstruction, with additional packages being deployed for higher level reconstruction (\eg, PID, pattern recognition, neutrino flavor identification). To optimise the reconstruction for atmospheric neutrinos, the standard accelerator reconstruction chain was modified in a number of targeted ways. First, a retraining on an atmospheric sample was done for the CNN and CVN networks used for vertexing and neutrino flavor determination algorithms, respectively. Then, improvements were brought to existing PID and energy reconstruction algorithms. Finally, specific for atmospheric event reconstruction, was the development of additional packages to determine particle direction (including that of neutrinos). 

The reconstruction efficiencies and resolutions results presented show a generally weak dependence on direction. Moreover, despite being a detector optimized for long-baseline reconstruction, it was demonstrated that the DUNE FD LArTPC technology delivers a demonstrably comparable (in overlap regions) and competitive (at lower and higher energy intervals) reconstruction performance for non-beam neutrino sources as well, and over a much wider energy range.

In addition, DUNE's unique capabilities at reconstructing the hadronic system with much lower detection thresholds than water-Cherenkov detectors allows for improved reconstructions of both neutrino energy and direction. At the multi-GeV scale, the DUNE resolution on the zenith angle is below \SI{10}{\degree} while the energy resolution is around \SI{15}{\percent}, a significant improvement with respect to the full-KM3NeT/ORCA expected resolutions~\cite{ORCA}. Moreover, it was shown in this paper that partially contained interactions can still be reconstructed with good performance in DUNE, meaning that DUNE's smaller size compared to that of other neutrino experiments is not expected to significantly impair reconstruction performance at those energies. All these will allow DUNE to obtain significant sensitivity to both the mass ordering and $\theta_{23}$ with atmospheric neutrino measurements.

It was also shown in this paper that DUNE is expected to perform well at detecting and measuring the energy and direction of neutrinos below \SI{2}{GeV}. At these neutrino energies, the sparser nature of the \si{Mton} scale detectors, such as KM3NeT/ORCA or IceCube, make this detection almost impossible. These interactions can be detected by \si{kton} scale detectors such as SuperKamiokande~\cite{SuperK}, but in this case, the reconstruction of the neutrino direction and energy relies almost-exclusively on the lepton measurement as most of the hadronic system is under the Cherenkov thresholds. But DUNE is able to accurately detect this hadronic system and leverage it for both energy and direction reconstruction improvements, leading to energy resolutions from \SIrange{10}{20}{\percent} for CC interactions, and improved angular resolution with respect to the lepton-only case. These characteristics makes DUNE a competitive experiment to obtain relevant sensitivities to the CP violation phase $\delta_{CP}$. Additionally, good, consistent performance across a wide range of energies and with $4\pi$ coverage should allow DUNE to probe several beyond-PMNS effects using atmospheric neutrinos.

This work lays the foundations for physics sensitivity studies with DUNE’s atmospheric neutrinos which will follow in separate future publications. The results show that atmospheric neutrinos can be reconstructed in the DUNE FDs with similar efficiency and resolutions as beam neutrinos (\numu, \nue), giving confidence that they can contribute and complement the success of the DUNE's physics program. The confirmation of increased reconstruction performance given the possibility of performing particle identification in the LArTPC, combined with the large size of the DUNE FDs using this technology, should give confidence to the neutrino community at large of robust future contributions from DUNE to the international atmospheric physics program. 


%% file: aknowlege.tex



\bigskip \hrule \bigskip

%
%
%
%
This document was prepared by DUNE collaboration using the resources of the Fermi National Accelerator Laboratory (Fermilab), a U.S. Department of Energy, Office of Science, Office of High Energy Physics HEP User Facility. Fermilab is managed by Fermi Forward Discovery Group, LLC, acting under Contract No. 89243024CSC000002.
%
%
This work was supported by
CNPq,
FAPERJ,
FAPEG, 
FAPESP and,
Funda\c{c}\~o Arauc\'{a}ria,             Brazil;
CFI, 
IPP and 
NSERC,                          Canada;
CERN;
ANID-FONDECYT,                  Chile;
M\v{S}MT,                       Czech Republic;
ERDF, FSE+,
Horizon Europe, 
MSCA and NextGenerationEU,      European Union;
CNRS/IN2P3 and
CEA,                            France;
PRISMA+,                        Germany;
INFN,                           Italy;
FCT,                            Portugal;
CERN-RO/CDI,                        Romania;
NRF,                            South Korea;
Generalitat Valenciana, 
Junta de Andaluc\'{i}a, 
MICINN, and 
Xunta de Galicia,               Spain;
SERI and 
SNSF,                           Switzerland;
T\"UB\.ITAK,                    Turkey;
The Royal Society and 
UKRI/STFC,                      United Kingdom;
DOE and 
NSF,                            United States of America.
%
%